\documentstyle[12pt,epsf]{article}
\begin{document}
\centerline{\large \bf Center Projection Vortices in 
Continuum Yang-Mills Theory}

\bigskip
\centerline{M.~Engelhardt\footnote{Supported by DFG under grant
number DFG En 415/1-1} and H.~Reinhardt\footnote{Supported in part
by DFG under grant number DFG Re 856/4-1} }
\vspace{0.2 true cm}
\centerline{\em Institut f\"ur Theoretische Physik, Universit\"at
T\"ubingen }
\centerline{\em D--72076 T\"ubingen, Germany}

\bigskip

\begin{abstract}
The maximal center gauge, combined with center projection, is a means to
associate Yang-Mills lattice gauge configurations with closed center vortex 
world-surfaces. This technique allows to study center vortex physics
in lattice gauge experiments. In the present work, the continuum
analogue of the maximal center gauge is constructed. This sheds new
light on the meaning of the procedure on the lattice and leads to
a sketch of an effective vortex theory in the continuum. Furthermore,
the manner in which center vortex configurations generate the
Pontryagin index is investigated. The Pontryagin index is built up
from self-intersections of the vortex world-surfaces, where it is
crucial that the surfaces be globally non-oriented.
\end{abstract}



\newpage

\section{Introduction}

The infrared sector of strong interaction physics is characterized by
nonperturbative phenomena such as confinement, the breaking of chiral
symmetry, and the $U_A (1)$ problem. These phenomena are believed to
be described by quantum chromodynamics; however, in order to understand
in more detail how they arise, it is helpful to isolate the particular
degrees of freedom of the theory which induce these phenomena.
Chiral symmetry breaking and the $U_A (1)$ problem can e.g. be successfully
explained in terms of instanton effects \cite{cdg}, whereas confinement 
eludes a description via instantons \cite{insnocf}.
\medskip

Viable explanations of confinement can be based on magnetic degrees of
freedom such as Abelian magnetic monopoles \cite{magmonv} or center 
vortices \cite{voride},\cite{spag}; considerable evidence in favor of such 
scenaria has accumulated in recent years from numerical lattice studies.
Magnetic monopoles emerge in Yang-Mills theory as gauge fixing defects in 
the so-called Abelian gauges \cite{tho81}, where the Cartan subgroup of
the gauge group under consideration is left unfixed. In all Abelian
gauges considered, the vacuum behaves as a dual superconductor 
\cite{digia}; in particular, lattice calculations performed in the 
maximal Abelian gauge \cite{cdommag},\cite{cdomma2} exhibit 
Abelian and monopole dominance for the string tension. On the other hand, 
lattice calculations performed in the so-called maximal center gauge 
\cite{deb97}, which will be discussed in detail in this work, have provided 
evidence in favor of the notion that confinement is induced by center 
vortices. The vortex content of gauge field configurations produces 
virtually the full string tension; conversely, the string tension disappears 
in the absence of vortices \cite{deb97}. This property of {\em center 
dominance} persists at finite temperatures \cite{tempv},\cite{tlang}, 
and the deconfinement phase transition can be understood as a transition 
from a percolating vortex phase to a phase in which the vortices cease to 
percolate.
\medskip

The aforementioned magnetic degrees of freedom inducing confinement can
furthermore be connected with the topological properties commonly
thought to be carried by instantons; they may therefore ultimately be 
suited to provide a unified picture of the different nonperturbative 
strong interaction phenomena mentioned further above. The magnetic monopoles
of the maximal Abelian gauge have been empirically found to be correlated 
with instantons in lattice calculations \cite{sugan}. In another Abelian
gauge, the Polyakov gauge, the Pontryagin index can be entirely
reconstructed from the magnetic monopole content of gauge field
configurations \cite{mmpoly}. On the other hand, one of the foci of the
present work is to elucidate the manner in which center vortex 
configurations generate the Pontryagin index; it will be shown that
the latter is produced by self-intersections of vortex world-surfaces,
where it is crucial that the surfaces be globally non-oriented. In
a recent lattice study \cite{bertle}, the orientability of the vortex
surfaces was studied, with the result that these surfaces are non-orientable
in the confining phase and have non-trivial genus. Furthermore, in an
ensemble devoid of vortices, chiral symmetry breaking disappears and all
configurations belong to the trivial topological sector \cite{forcrand}.
\medskip

For the abovementioned lattice studies of vortex properties to become
possible, techniques had to be developed which allow to associate a
given lattice gauge configuration with a configuration of vortex 
world-surfaces. Such techniques have only become available in relatively
recent times; this has revived interest in the vortex picture, which 
received rather little attention after some early efforts following its 
inception \cite{voride},\cite{spag},\cite{vortvan}. The techniques
referred to here are the maximal center gauge and center projection
\cite{deb97},\cite{deb97aug} already mentioned further above.
While not all questions concerning the numerical stability of this
procedure and its ability to faithfully represent thick center fluxes
present in full gauge configurations have been settled \cite{tomuns},
it has facilitated a number of observations concerning vortex physics
in addition to those already mentioned further above. E.g., the density of
the center vortices obtained via the maximal center gauge and center
projection displays the proper scaling behavior required under the
renormalization group and, consequently, these vortices can be considered 
as physical objects \cite{kurt} (note erratum in \cite{tempv}), 
see also \cite{giedt}. Also binary correlations of vortex intersection
points found on a given space-time plane display such scaling \cite{corr}.
Furthermore, center vortices can account for the Casimir scaling
behavior of the adjoint string tension if one associates a finite
physical thickness with the vortex world-surfaces \cite{adj1},\cite{adj2}.
In an indirect version of the maximal center gauge, which proceeds via
an intermediate maximal Abelian gauge fixing, both Abelian magnetic
monopoles and center vortices were detected \cite{deb97aug}; the monopoles
were found to be located on center vortices like beads on a string.
\medskip

One of the foremost aims of this work is to develop a continuum
formulation for the center gauge and center projection procedure. This
will also shed some additional light on issues concerning the procedure
on the lattice. It should be clear that the attempt to construct an
analogue of the maximal center gauge in the continuum encounters some 
conceptual obstacles not found in continuum formulations of Abelian
gauges, which are quite straightforward. These obstacles are related to
the fact that the continuum theory is constructed in terms of a gauge 
potential, defined in the adjoint representation of the gauge group, as 
opposed to the lattice formulation in terms of group elements on lattice 
links. The continuum gauge fields are thus insensitive to the center of the
gauge group. It is a priori not clear how to extend the successful
concept of center projection to the continuum theory. Nevertheless,
it will prove possible in this work to derive a continuum analogue
of the lattice procedure.
\medskip

In detail, the plan of this paper is as follows. In section \ref{mcglatsec},
the maximal center gauge and the emergence of center vortices after center
projection is carefully analyzed on the lattice. The effect of this gauge
on sample gauge fields of the vortex type is studied empirically; 
subsequently, the gauge fixing procedure on the lattice is reformulated in 
a manner which explicitly separates the $Z(N)$ center and $SU(N)/Z(N)$ coset 
degrees of freedom. The continuum limit of center and coset lattice 
configurations is considered in section \ref{tcontlim}, along with the
gauge transformations which arise in the course of maximal center gauge
fixing. Explicit continuum representations of center vortices are given.
This allows the construction of the continuum analogue of the maximal 
center gauge, the properties of which are then discussed extensively,
including alternative formulations which allow to extract more general 
objects of the vortex type; for reasons mentioned further above, vortex 
surfaces made up of patches of different orientation are of particular 
interest, as well as thickened vortices. In section \ref{effthsec},
the implementation of the maximal center gauge in the continuum path 
integral is considered, and a sketch of an effective vortex theory is given.
Finally, the Pontryagin index generated by center vortex configurations
is studied in section \ref{topsect}. The concluding section \ref{outlsec}
discusses some open issues to be settled in subsequent work. Some 
mathematical details are presented in the appendices. 
\newpage

\section{Maximal Center Gauge and Center Vortices}
\label{mcglatsec}
\subsection{The Maximal Center Gauge}
\medskip
Maximal center gauge is a gauge fixing condition defined in the lattice
formulation of $SU(N)$ Yang-Mills theory. It prescribes that the gauge 
freedom
\begin{equation}
U_{\mu } (x) \rightarrow 
U_{\mu }^{V} (x) = V (x + e_{\mu } ) U_{\mu } (x) V^{\dagger } (x)
\label{latgtra}
\end{equation}
be used to transform the link variables $U_{\mu } (x)$ of a given lattice
configuration as close as possible to center elements of the gauge
group, in a sense still to be specified. In fact, there exist many
different specifications, or realizations, of the general idea
formulated above. One possible explicit gauge fixing condition,
first introduced in \cite{deb97aug} for the case of $SU(2)$ 
color\footnote{Generalizations and applications to $SU(3)$ color
were discussed in \cite{giedt},\cite{montero}.} as the 
``Direct Maximal Center Gauge'', is
\begin{equation}
\max_{V} \sum_{x,\mu } \left| \mbox{tr} \, U_{\mu }^{V} (x) \right|^{2}
\label{mcg}
\end{equation}
for arbitrary local $SU(2)$ gauge transformations $V$.
\medskip

In order to make this more transparent, represent the $SU(2)$ link 
variables as
\begin{equation}
U = c_0 + i \vec{c} \vec{\tau }
\end{equation}
where $\tau_{i} $ denotes the Pauli matrices and the parameters 
$c_0, c_i, i = 1,2,3$, fulfil the constraint
\begin{equation}
c_{0}^{2} + \vec{c} \,^{2} = 1
\label{unit2}
\end{equation}
to guarantee unitarity of the links. Note that (\ref{unit2}) defines a 
three-sphere in the space of parameters $c_0, c_i $. In the following,
the set of group elements with $c_0 > 0$ will sometimes be referred to
as the ``northern hemisphere'' of the $SU(2)$ group, and the set of
group elements with $c_0 < 0$ as the ``southern hemisphere''. The
poles of the sphere, $c_0 = \pm 1$, correspond to the center $Z(2)$
of the $SU(2)$ group. Correspondingly, the gauge condition (\ref{mcg})
reduces to
\begin{equation}
\max_{V} \sum_{x,\mu } \left( c^{V}_{0, \mu } (x) \right)^{2}
\end{equation}
which can be straightforwardly visualized in terms of a bias of the link
variables towards the poles of the $SU(2)$ three-sphere.
\medskip

An important property of the condition (\ref{mcg}) is the fact that it
leaves the center part of the gauge freedom, i.e. (for the general
case of a $SU(N)$ gauge group) the $Z(N)$ part, unfixed. 
There is no bias distinguishing between different center
elements due to the absolute value prescription in (\ref{mcg}).
This is analogous to the so-called Abelian gauges \cite{tho81}, which 
leave the Abelian part of the gauge freedom, i.e. the $[U(1)]^{N-1}$ 
part, unfixed.
\medskip

The specific condition (\ref{mcg}) is only one of many ways to bias
link variables towards the center of the gauge group. In fact, any
condition of the type
\begin{equation}
\max_{V} \sum_{x,\mu } \bar{g} \left( 
\left| \mbox{tr} \, U_{\mu }^{V} (x) \right| \right)
\label{mcgg}
\end{equation}
with a monotonously rising function $\bar{g} $ implements the general
idea of maximal center gauge fixing.
\medskip

After transforming a lattice gauge configuration to the maximal center
gauge, one can carry out a second step, called {\em center projection},
which serves to define the $Z(N)$ vortex content of the configuration.
Center projection corresponds to mapping every link variable $U$ onto 
the center element closest to $U$. In the case of the $SU(2)$ gauge
group, this simply means mapping
\begin{equation}
U \rightarrow \mbox{sign tr } U \ ,
\label{cpgen}
\end{equation}
i.e.
\begin{equation}
c_0 \rightarrow \mbox{sign } c_0 \ .
\end{equation}
In other words, all link variables located on the northern hemisphere are
mapped onto the north pole of the $SU(2)$ group, all links on the 
southern hemisphere onto the south pole. For general $SU(N)$ group,
center projection leaves one with a $Z(N)$ lattice, which in the canonical
fashion can be associated with $(D-2)$-dimensional vortex surfaces
(where $D$ denotes the space-time dimension) on the dual lattice:
If any plaquette on the $Z(N)$ lattice evaluates to a nontrivial
center element, it is considered to be pierced by a vortex carrying
the appropriate (quantized) flux; if it takes the trivial value unity,
no vortex pierces the plaquette. The vortices defined in this way
represent $(D-2)$-dimensional surfaces in $D$ space-time dimensions.
The center flux carried by the vortices is continuous; there are no sources
or sinks for this flux. In the case of a $Z(2)$ lattice, this corresponds
to the vortex surfaces being closed. In the general $Z(N)$ case, the 
surfaces may branch, while respecting flux conservation, since more than
one nontrivial center flux exists. However, precisely due to flux
conservation, one may think also of such branched surfaces in the
$Z(N)$ case as being composed of simple closed surfaces which
in general share whole surface patches in space-time.
\medskip

\subsection{Center vortices on the lattice}
\label{examse}
For the purpose of illustrating the effect of the center gauge
condition, consider the following sample continuum $SU(2)$ gauge field
configurations on a two-dimensional plane. Let the plane be parametrized
in polar coordinates $r,\varphi $ and define the gauge field 
as\footnote{Gauge fields in the following are decomposed using
antisymmetric generators $T_a $ of the gauge group satisfying
$[T_a ,T_b ] = f_{abc} T_c $, with $f_{abc} $ denoting the structure
coefficients, and the normalization Tr$T_a T_b = -\delta_{ab} /2$.
For $SU(2)$, $T_a = -i\tau_{a} /2 $, where $\tau_{a} $, $a=1,2,3$,
denotes the Pauli matrices.}
\begin{equation}
a_{\varphi } = \pm \frac{1}{r} T_3 \ , \ \ \ a_r =0
\label{thinv}
\end{equation}
This configuration, and the ones to follow below, can be regarded as 
slices of configurations defined in more space-time dimensions, and 
which, for the present purposes, can be considered constant in the 
additional dimensions. Also the Lorentz components of the gauge field 
associated with those additional directions should be considered to vanish.
The above configuration is a particular example of what in the 
following will be called a {\em thin vortex} (positioned at the origin 
of the plane). A general representation of thin vortices will be given
in eq. (\ref{arbthv}). The defining property of a thin vortex is that
any Wilson loop which encircles the vortex, i.e. in the present case, 
the origin, takes a nontrivial value in the center of the gauge group.
For the case of $SU(2)$ color, this corresponds to the value $(-1)$. 
On the other hand, any other Wilson loop, not encircling the vortex, takes 
the value $(+1)$. In other words, the vortex contributes a nontrivial center 
element of the gauge group to any Wilson loop it pierces. Correspondingly, 
the field strength is singular and localized at the position of the vortex. 
Note that both choices of sign in (\ref{thinv}) lead to these properties. 
Thus, one can distinguish two different orientations of vortices, 
associated with the direction of (magnetic) flux they describe (flux which 
is dual to Wilson loops will sometimes be referred to as ``magnetic'' in 
the following). On the other hand, also (odd) integral multiples of the 
configurations (\ref{thinv}) satisfy the defining properties of center
vortices. However, such configurations will not be considered as distinct
types of vortices, but simply as superpositions of vortices of the smallest
Abelian magnetic flux required to induce the center element in question
in a Wilson loop.
\medskip

Consider furthermore the more general configuration
\begin{equation}
A_{\varphi } = \pm \frac{f(r)}{r} T_3 \ , \ \ \ A_r =0
\label{thickv}
\end{equation}
with $f(0)=0$ and $f(\infty ) =1$. This configuration represents
a {\em thick vortex}; in order for a Wilson loop to take the value
$(-1)$, it must circumscribe the vortex at a sufficient distance,
the thickness of the vortex, determined by the form of the
{\em profile function} $f(r)$. Also the field strength or the flux 
associated with a thick vortex are spread out as compared to the
singular localized field strength of a thin vortex (\ref{thinv}).
For certain purposes, such as calculating field strengths, it may be 
necessary to specify more precisely what one means by the singularity 
of the thin vortex (\ref{thinv}) at the origin; this becomes relevant 
e.g. in Appendix \ref{appc}. In such a case, the thin vortex should always
be viewed as the limit in which $f(r)\rightarrow 1$ for $r\neq 0$,
but still $f(r=0)=0$ in (\ref{thickv}).
\medskip

These configurations can be put on a two-dimensional lattice in the
canonical fashion by associating the path-ordered exponential
\begin{equation}
U_{\mu } (x) = {\cal P} \exp \left( -\int_{P} A_{\mu } (z) dz_{\mu } \right)
\end{equation}
with a link $U_{\mu } (x) $ describing a path $P$ originating from the
point $x$ and running in the $\mu $-direction. Note that path-ordering
${\cal P} $ is not actually necessary for the Abelian configurations 
(\ref{thinv}) and (\ref{thickv}). This procedure of putting the continuum 
configurations on the lattice preserves the values of Wilson loops taken 
in the corresponding configurations, as described above.
\medskip

\begin{figure}[ht]

\vspace{-9cm}

\centerline{
\epsfysize=14.5cm
\epsffile{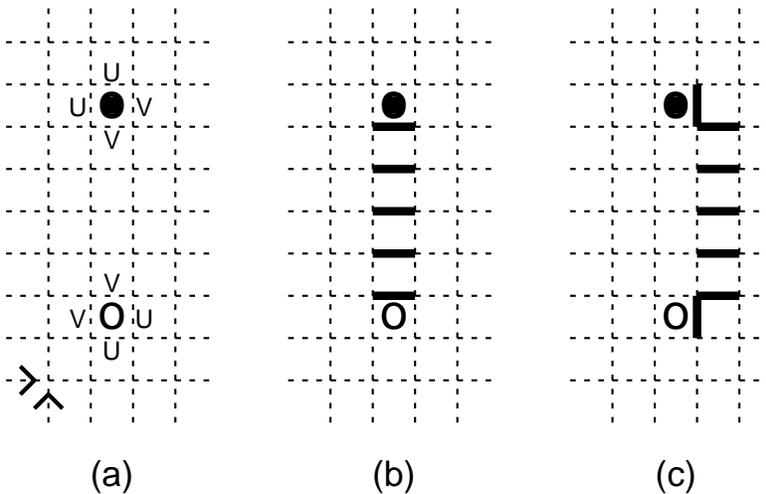}
\vspace{0.7cm}
}
\caption{Two thin SU(2) vortices, cf. eq. (\ref{thinv}), placed on
a space-time lattice (a), where the arrows in the lower left-hand corner
specify the direction of the line integrals defining horizontal and
vertical links, respectively. The open and filled circles denote the 
locations of the vortices, the difference in the symbols signifying that 
the vortices have opposite orientation. In (a), all lattice links deviate
from unity, but all plaquettes multiply to (+1), except the plaquettes
pierced by vortices, which multiply to (-1). The links making up
those latter plaquettes take the values $U=\exp (T_3 \pi/2)$ and
$V=U^{\dagger } $ (up to small corrections from the respectively
other vortex). After maximal center gauge fixing (b), all lattice links
take the value (+1), except the ones displayed in fat print, which
take the value (-1). There are alternative gauge fixing images, such
as (c), for which the center gauge fixing functional
is degenerate, and which are related to the configuration (b) by
gauge transformations purely from the center of the gauge group.}
\label{fig1}
\end{figure}

In practice, it is desirable to treat a lattice of finite extent with
periodic boundary conditions, i.e. a lattice torus. In order to fulfil 
such boundary conditions, it is necessary to have zero net flux piercing 
the plane under consideration\footnote{Consider a Wilson loop running 
around the edge of the lattice. Due to the periodic boundary conditions and
the Abelian character of the configurations treated here, contributions
from opposite sides of the loop cancel and the Wilson loop must take
the value $(+1)$.}. For this reason, in the following, linear
superpositions of pairs of (thin or thick) vortices of opposite
orientation will be examined, cf. Fig.~\ref{fig1} (a). Due to the Abelian
character of the configurations defined above, the phases
contributed to Wilson loops by single vortices in a many-vortex 
superposition simply add, without interference from path-ordering
effects. Furthermore, in calculating the lattice links from the 
gauge fields, also the contributions from periodic images of the 
vortices must be taken into account\footnote{In practice, for better 
convergence of this sum, the authors covered the plane with positively 
and negatively oriented vortices in checkerboard fashion. Therefore, 
the lattice torus actually used in numerical computation contained 
four vortices as opposed to the two displayed in Fig.~\ref{fig1}. This is 
merely a point of technical convenience.}.
\medskip

Having put the gauge configurations on the lattice, the question 
can be addressed what effect the center gauge fixing procedure has on 
these configurations. For the case of two thin vortices, fixing the 
configuration depicted in Fig.~\ref{fig1} (a) to maximal center gauge 
can be carried out trivially without any recourse to numerical computation. 
All one needs are the following two observations: The configuration shown 
in Fig.~\ref{fig1} (b) is gauge equivalent to the one shown in 
Fig.~\ref{fig1} (a) since all Wilson loops take the same values in the 
two configurations. At the same time, in Fig.~\ref{fig1} (b) all links 
take values in the center of the gauge group. Therefore, this configuration
realizes the maximal possible value of the gauge fixing functional
(\ref{mcgg}). Thus, the configuration shown in Fig.~\ref{fig1} (b) is 
precisely a center gauge fixing image of the configuration shown in 
Fig.~\ref{fig1} (a). It is described by a string of links taking a value 
corresponding to a nontrivial center element extending from the position of 
one thin vortex to the position of the other. This type of configuration,
i.e. strings of links associated with a nontrivial center element
of the gauge group, will in the following be termed a configuration
of {\em ideal vortices}. It is, as just seen, gauge equivalent to a
configuration of {\em thin vortices} at the positions where the strings
end.
\medskip

It should be noted that there exist many other degenerate maxima of the 
gauge fixing functional on the same gauge orbit, namely all deformations
of the string of links, cf. Fig.~\ref{fig1} (c). These are all images of 
the configuration shown in Fig.~\ref{fig1} (b) under gauge transformations 
which are elements of the center of the gauge group. This is not
surprising; by construction, the gauge fixing condition (\ref{mcgg}) is
invariant under such transformations (e.g. in the $SU(2)$ case,
it is insensitive to the sign of the link variables). In other words,
it leaves the center of the gauge group unfixed. As a consequence,
the precise trajectory of the string of nontrivial links describing
ideal vortices is arbitrary; only its two endpoints (corresponding
here to the positions of the two thin vortices) are fixed, and they
embody all gauge-invariant information.
\medskip

Note that in three-dimensional space-time, the sets of nontrivial
links describing ideal vortices generically form two-dimensional
sheets, whereas in four-dimensional space-time, they form
three-dimensional volumes. The boundaries of the sheets or volumes,
respectively, describe the locations of the thin vortices to which
the ideal vortex configuration is gauge equivalent.
\medskip

As a further aside, note that center projection leaves the center
gauge fixed configuration (Fig.~\ref{fig1} (b)) unchanged and thus does 
not truncate any physical information contained in it. By contrast,
center projecting the original configuration (Fig.~\ref{fig1} (a)) 
before gauge fixing completely removes the vortices and yields a trivial
lattice with all links at unity.
\medskip

\begin{figure}[ht]

\vspace{-14cm}

\centerline{
\epsfysize=21cm
\epsffile{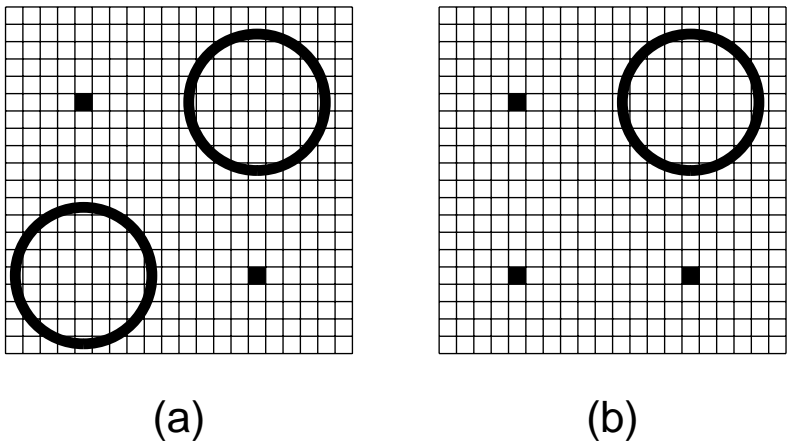}
}
\caption{Configurations containing thin vortices and thick vortices,
the flux of the latter being concentrated on the displayed rings.
These configurations can be transformed into the maximal center
gauge numerically (see text).}
\label{fig2}
\end{figure}

Continuing to the case of a thick vortex, it is clear that this case 
cannot be solved in a similarly trivial fashion. Here, the latticized
gauge configurations can be brought into the maximal center gauge 
numerically by maximizing the gauge functional (\ref{mcgg}) under the
action of arbitrary $SU(2)$ gauge transformations on the sites of the 
lattice. As described above, lattice tori containing four vortices
were used. Two different cases were examined: Two thick vortices
of one orientation together with two thin vortices of the other
orientation (cf. Fig.~\ref{fig2} (a)), and one thick vortex together 
with three thin vortices, again such that the total flux piercing the 
plane vanishes (cf. Fig.~\ref{fig2} (b)). In order to make the new 
effects arising in the case of a thick vortex more visible, a Fermi 
distribution
\begin{equation}
f(r)=1-\frac{1}{1+\exp (5(r-4l)/l)}
\label{ringv}
\end{equation}
where $l$ denotes the lattice spacing,
was chosen for the profile function $f(r)$ (cf. eq. (\ref{thickv})).
This not only spreads out the flux of the vortex but actually
localizes it on a ring of radius $4l$ concentric to the core 
of the vortex.
\medskip

The gauge equivalent configurations which maximize
the center gauge fixing functional (\ref{mcgg}) will contain a certain 
set of links which lie in the ``southern hemisphere'' of the $SU(2)$
group, i.e. nearer to the center element $(-1)$, whereas the
other links remain closer to the center element $(+1)$. Center
projection then associates a collection of ideal vortices, or,
gauge-equivalently, thin vortices, with the original configuration.
Carrying out this procedure numerically for the configurations depicted
in Fig.~\ref{fig2}, one obtains thin vortices which approximate
the locations of the original thick vortices; however, the precise
location of the thin vortex within the profile of the thick vortex turns 
out to depend on the precise realization of the center gauge via the 
arbitrary function $\bar{g} $ in the gauge fixing functional (\ref{mcgg}). 
Specifically, for the configuration containing one thick vortex, 
Fig.~\ref{fig2} (b), it turns out that for the gauge fixing function 
$\bar{g} (t)=t^2 $, the thin vortex is located concentrically with the 
thick vortex, whereas for $\bar{g} (t)= 1-\theta (2-t) (1-t^2 /4)^2 $, the 
thin vortex is located one lattice spacing away from the center of the 
thick vortex. This latter effect becomes more pronounced if one uses the 
configuration with two thick vortices, Fig.~\ref{fig2} (a), again with 
the gauge fixing function $\bar{g} (t)= 1-\theta (2-t) (1-t^2 /4)^2 $.
There, the thin vortex is located four lattice spacings away
from the center of the thick vortex, i.e. on the ring of flux
depicted in Fig.~\ref{fig2} (a).
\medskip

The conclusion reached from these examples is therefore that
the maximal center gauge does yield thin vortices which roughly 
approximate the locations of thick vortices in gauge field configurations.
This corroborates the findings of \cite{deb97},\cite{montero}; note
however recent indications of considerable instability in the numerical 
gauge-fixing procedure \cite{tomuns}. Furthermore, the precise 
coordinates of the thin center projection vortices turn out to be 
gauge-dependent, via the gauge fixing function $\bar{g} $. Also, 
comparing the results obtained with the two different configurations 
depicted in Fig.~\ref{fig2}, the precise location of a thin vortex 
approximating a given thick vortex evidently also depends on the structure 
of other vortices in the vicinity. Therefore, it seems that the
center gauge fixing procedure introduces nonlocal interactions into
the center projection vortex effective theory independent of any
reference to the Yang-Mills dynamics which ultimately governs the
original thick vortex configurations.
\medskip
 
\subsection{Separating center and coset degrees of freedom}
\label{sepsec}
In the attempt to formulate an analogue of the center gauge and center 
projection procedure in continuum Yang-Mills theory, one obviously 
encounters a conceptual obstacle: Continuum Yang-Mills theory is
described in terms of a gauge potential $A_{\mu}^{a} (x)$ which is 
defined in the adjoint representation of the gauge group. Thus, in the 
continuum, the gauge symmetry of Yang-Mills theory is only a local 
$SU(N)/Z(N)$, i.e. a coset, symmetry. Related considerations concerning
this issue in the Schr\"odinger picture can be found in \cite{haage}. The 
first step in dealing with this problem is to carefully separate center 
and coset degrees of freedom in the lattice gauge fixing procedure
discussed above. The treatment below will for simplicity mainly
concentrate on the case of an $SU(2)$ gauge group; however, the
generalization to any $SU(N)$ group is conceptually, if not
necessarily notationally, straightforward. One only has to keep
in mind that higher $SU(N)$ groups have more center elements, and
that therefore there are more species of vortices, corresponding
to the different possible fluxes. The only qualitative difference
in terms of the topology of vortex configurations is that, for higher
gauge groups (and three or more space-time dimensions), vortices may 
branch into several vortices with different center fluxes, as long as 
flux conservation is respected at branchings.
\medskip

To have a slightly more precise terminology at hand in the discussion
below, it is useful to decompose the link variables $U_{\mu } (x)$
describing a $SU(2)$ lattice configuration into a center part
\begin{equation}
Z_{\mu } (x) = \mbox{sign tr} \, U_{\mu } (x)
\end{equation}
and a coset part $U^{\prime }_{\mu } (x)$ defined by 
\begin{equation}
U_{\mu } (x) = Z_{\mu } (x) U^{\prime }_{\mu } (x) \ .
\end{equation}
The lattice described by $Z_{\mu } (x) $ will be referred to as the
{\em center part} of the lattice configuration $U_{\mu } (x)$,
whereas the lattice described by $U^{\prime }_{\mu } (x) $ will be referred
to as the {\em coset part}. Note that the center gauge fixing functional
(\ref{mcgg}) depends only on the coset part of a configuration; center 
projection simply means mapping $U_{\mu } (x)$ onto $Z_{\mu } (x) $.
\medskip

Furthermore, below the terminology {\em coset transformation} will be
used to characterize gauge transformations, cf. eq. (\ref{latgtra}), which 
change only the coset part of a lattice configuration. Note that this does 
not imply that the corresponding unitary transformation matrices acting at 
lattice sites are necessarily from the northern hemisphere of the $SU(2)$ 
group. Note furthermore that the set of possible coset transformations 
depends on the initial gauge configuration and that in general not all 
members of this set can be continuously deformed to the identity while 
remaining in the set. Also, smooth gauge transformations $G$ in the 
continuum formulation of Yang-Mills theory,
\begin{equation}
A_{\mu } \longrightarrow A_{\mu }^{G} = 
G(A_{\mu } +\partial_{\mu } ) G^{\dagger }
\end{equation}
are by construction coset gauge transformations.
\medskip

Clearly, if one puts a smooth continuum gauge configuration $A_{\mu } (x)$
onto a sufficiently fine lattice, the associated center part is trivial,
$Z_{\mu } (x) =1$. However, even in such a case, gauge fixing in general,
and the center gauge fixing procedure in particular, may induce a 
nontrivial center part in the gauge fixed lattice; explicit examples were 
given in the previous section. In other words, some links in the gauge 
fixed lattice end up in the southern hemisphere, i.e. closer to the 
nontrivial center element $(-1)$ than to the trivial element $(+1)$; 
ideal vortices are introduced.
\medskip

In the $SU(2)$ lattice description of the center gauge fixing procedure,
one allows arbitrary $SU(2)$ gauge transformations when maximizing the
gauge fixing functional (\ref{mcgg}). This procedure can be broken down 
into smaller steps as follows. 
\begin{enumerate}
\item
Start from a configuration with a trivial center part, i.e. with all links
in the northern hemisphere, such as one obtains by putting any smooth 
continuum gauge configuration onto a sufficiently fine lattice. Apply a 
gauge transformation which rotates a certain arbitrary, but fixed, set of 
links to the southern hemisphere, i.e. such that the center part
of the transformed lattice displays the nontrivial value $(-1)$ on that 
set of links\footnote{As long as the original links are sufficiently 
close to unity (which one can always achieve by choosing the lattice 
spacing accordingly), this transformation can e.g. be effected by applying 
the group element $V=\exp (2\pi T_3 /3)$ to the site from which the links 
in question emanate and the group element $V^{\dagger } $ to the site at 
which they end, i.e. $U\rightarrow VUV$, where $U$ denotes the original
link variable associated with the link under consideration. One can also 
exchange $V$ and $V^{\dagger } $, which may become necessary if two links 
in question share a site.}.
\item
Maximize the gauge fixing functional using only coset transformations, 
i.e. keeping the center part of the configuration invariant. In other
words, links are not allowed to switch ``hemispheres'' of the $SU(2)$
group. Note that this second step operates purely on the coset part
of the lattice configuration obtained after step 1.) above; not only 
are only coset transformations allowed, but also the gauge fixing 
functional (\ref{mcgg}) depends exclusively on the coset part of the 
link variables.
\item
Finally, this two step procedure must be repeated for all possible 
first step choices of the center part of the transformed lattice, 
in order to cover all possible gauge transformations contained in the 
full $SU(2)$ gauge group. For each of these choices, corresponding
to a certain ideal vortex configuration, one obtains in the
associated second step a constrained maximum of the gauge fixing functional 
(\ref{mcgg}); from all these maxima, one must choose the highest one in 
order to obtain the gauge-fixed configuration reached under the action of
the full $SU(2)$ gauge group.
\end{enumerate}
\medskip

Slightly rephrased, the procedure described above splits the process of 
center gauge fixing under the $SU(2)$ gauge group up into the following 
parts: One introduces, by a suitable gauge transformation, into an initially 
smooth gauge configuration ideal vortices at fixed, but arbitrary, 
locations.  This step transfers information from the coset part into
the center part of the configuration. From this point onwards, however,
center and coset parts remain strictly separated and subsequent gauge
transformations operate only on the coset part. The latter is transformed, 
by using purely coset transformations, to the (constrained) maximum of 
the (purely coset-dependent) gauge fixing functional (\ref{mcgg}). 
The algorithm must be repeated for all possible choices of ideal vortex 
configurations. The choice which leads to the global maximum of the gauge 
fixing functional determines the center gauge and center projection image 
of the original configuration.
\medskip

Formally, one can summarize this as follows. The center gauge fixing
condition, cf. eq. (\ref{mcgg}), corresponds to the two-fold maximization 
problem
\begin{equation}
\max_{Z} \max_{G} \sum_{x,\mu } \bar{g} \left( \left| \mbox{tr} \left( 
\left( U_{\mu } (x)^{V[Z]} \right)^{\prime \, G} \right) \right| \right)
\label{sepgf}
\end{equation}
where $U_{\mu } (x) $ is the gauge configuration one starts with and
$Z$ denotes the center part of the transformed lattice chosen in step 1.) 
above, realized by the gauge transformation $V[Z]$ as specified
there. The prime emphasizes explicitly that only the coset part of
the resulting configuration is considered further, although this is
already guaranteed by the absolute value prescription. In other words,
the configurations $(U^V)^{\prime } $ are stripped of all ideal vortices. 
Finally, $G$ denotes all coset gauge transformations considered in 
step 2.) above.
\medskip

\section{Taking the continuum limit}
\label{tcontlim}

\subsection{Continuum limit of lattice configurations}

As the first step in formulating a continuum version of the maximal
center gauge, it is necessary to define the continuum analogues of the 
various lattice configurations considered above.
\medskip

\subsubsection{Continuum limit of the coset part}

The coset part of a lattice configuration can be directly
associated with a regular continuum gauge field constructed such as
to reproduce all links of the coset configuration if put on the
lattice. One could e.g. achieve this construction by giving an
``inverse blocking'' prescription, successively replacing coarser
by finer lattices while dividing up the phases carried by the coarse
links onto the fine links such as to achieve a smooth continuum
limit. In this limit, it will be possible to expand an exponential
representation of the link variables
\begin{equation}
U^{\prime }_{\mu  } (x) = \exp (-l A_{\mu }^{\prime } (x) )
\label{cosexp}
\end{equation}
into powers of the lattice spacing $l$ and regard $A_{\mu }^{\prime } (x) $
as the continuum gauge field. As a side remark, note that if one wished 
to apply such an ``inverse blocking'' procedure to the center part of a 
configuration, the corresponding prescription would have to be radically 
different in order to preserve the center, or ideal vortex, character of 
the configuration: If a nontrivial center link of a coarse lattice is
divided up into two links of a finer lattice, the whole nontrivial
center phase would have to be associated with one of the two links,
whereas the other link would be associated with unity. The phase is
not divided up among the two links, in contrast to the coset case,
and the continuum limit is singular. In particular, expanding an
exponential representation analogous to (\ref{cosexp}) is by
construction never possible for the center part of a configuration.
Due to these difficulties, a different approach will have to be taken 
when discussing the continuum limit of the center part below.
\medskip

Before proceeding, it is worth noting that the gauge freedom associated with
the continuum gauge field $A_{\mu }^{\prime } $ in (\ref{cosexp}) matches
the one associated with the coset lattice $U_{\mu }^{\prime } $ in 
step 2.) of the center gauge fixing procedure as described in 
section \ref{sepsec}; namely, it consists of all possible coset 
transformations. Thus, in describing a coset lattice in the continuum limit
by means of a gauge field $A_{\mu }^{\prime } $ as in (\ref{cosexp}),
one is not introducing any restriction on the residual gauge freedom
allowed in step 2.) of the center gauge fixing procedure. The additional 
gauge freedom implied by the full $SU(N)$ symmetry present in lattice 
Yang-Mills theory has been explicitly separated off into the choice of 
ideal vortex configuration in step 1.).
\medskip

\subsubsection{Continuum limit of the center part}
\label{clotcp}

The center part of a configuration contains exclusively ideal
vortices, described by sets of links taking a nontrivial value from
the center of the gauge group. At this point, it is useful
to give an alternative characterization of such a configuration
which does not make explicit reference to a space-time lattice. 
This can be achieved via the effect the configuration has on
arbitrary Wilson lines. Consider therefore calculating an arbitrary 
Wilson line; as one travels along the line, gathering up phases from 
the different links making up the path, the Wilson line either stays 
constant or picks up a phase corresponding to a nontrivial element from
the center of the gauge group on a length of one lattice 
spacing $l$. Expressed through the vortex content, this latter change 
happens whenever the Wilson line intersects a string, surface, or volume 
of nontrivial links (depending on the space-time dimension) describing 
an ideal vortex configuration. As the lattice spacing is taken to zero, 
these strings, surfaces, or volumes become infinitely thin.
Therefore, the continuum analogue of an ideal vortex configuration
consists in specifying strings, surfaces, or volumes $\Sigma $
(depending on the space-time dimension) which, when intersected by
a Wilson line, contribute a center element factor to the latter.
\medskip

\subsection{Singular gauge transformations}
\label{singgt}
In the previous sections, the effect of the maximal center gauge on 
vortex gauge fields originally defined in the continuum theory was
investigated by first putting the Yang-Mills field on the lattice, 
subsequently implementing the maximal center gauge condition (which, 
so far, is defined only on the lattice) and afterwards considering 
again the continuum limit. One would, however, prefer not to use the 
detour via the lattice, but rather perform the maximal center gauge
fixing directly in the continuum. Particular care is needed in 
defining the analogue of step 1.) of the maximal center gauge fixing
procedure as laid out in section \ref{sepsec}.
It should be clear from the previous discussion that this step
in the continuum will introduce singularities 
(the ideal vortices) into the initially smooth gauge configurations. 
As already partly indicated above, these singularities are to be 
defined via their lattice origin, taking care in particular that 
gauge invariant quantities are unaffected by gauge transformations, 
even when the latter have a singular continuum limit. Keeping this 
methodology in mind, it will be possible to paraphrase the center 
gauge fixing procedure, as described step by step in section \ref{sepsec},
directly in the continuum.
\medskip

Consider a continuum Abelian gauge transformation $V(k,\Sigma ,x)$ specified
as follows. Let $V$ be continuous everywhere except at hypersurfaces 
$\Sigma $ of dimension $D-1$ if $D$ is the dimension of space-time.
At such hypersurfaces, $V$ shall have the following property. Let
$t$ denote the local coordinate perpendicular to $\Sigma $, with
$t=0$ on $\Sigma $, and let $x_{\perp } $ denote the other local
space-time coordinates. Then
\begin{eqnarray}
V(k,\Sigma , x_{\perp } , t\! =\! \epsilon ) 
V^{\dagger } (k,\Sigma , x_{\perp } , t\! =\! -\epsilon ) &=& 
Z(k) \label{vdef1} \\
V(\Sigma , x_{\perp } , t\! =\! -\epsilon ) \partial_{\mu }
V^{\dagger } (\Sigma , x_{\perp } , t\! =\! -\epsilon ) &=&
V(\Sigma , x_{\perp } , t\! =\! \epsilon ) \partial_{\mu }
V^{\dagger } (\Sigma , x_{\perp } , t\! =\! \epsilon )
\nonumber
\end{eqnarray}
where $\epsilon \rightarrow 0 $ and $k$ labels the $N-1$ different 
nontrivial center elements $Z(k)$ of the $SU(N)$ gauge group under
consideration. In other words, $V$ jumps by a center
phase $Z(k)$ at hypersurfaces $\Sigma $. An explicit realization of such
a gauge transformation $V$ will be given in the next section.
\medskip

Consider applying a gauge transformation such as characterized above
to an initial smooth field configuration $A$,
\begin{equation}
A\rightarrow A^V = VAV^{\dagger } + V\partial V^{\dagger }
\end{equation}
The first term on the right hand side still represents a smooth field
everywhere except possibly on the edge $\partial \Sigma $ of $\Sigma $, 
since it is quadratic in $V$ and hence insensitive to the center phase
$V$ picks up at hypersurfaces $\Sigma $. More care must be taken
in properly defining the singularity in the second term. For this
purpose, consider a Wilson line describing a straight-line path $C$
from $(x_{\perp } , t=-\epsilon )$ to $(x_{\perp } , t=\epsilon )$,
with $x_{\perp } $ and $t$ defined as above, and infinitesimal $\epsilon $.
Since the Wilson line transforms as
\begin{equation}
W [A] (C) \rightarrow W [A^V ] (C) =
V (x_{\perp } , t=\epsilon ) W [A] (C) 
V^{\dagger } (x_{\perp } , t=-\epsilon )
\end{equation}
and furthermore $W[A] (C) = 1+O(\epsilon )$, one has
\begin{equation}
W [A^V ] (C) = Z(k) + O(\epsilon )
\end{equation}
Therefore, the singularities of $A^V $ at the hypersurfaces $\Sigma $ 
describe nothing but an ideal vortex configuration $ {\cal A} (\Sigma )$. 
On the other hand, the term $V\partial V^{\dagger } $ is a pure gauge. 
Therefore, the regular part of $V\partial V^{\dagger } $ must compensate 
the ideal vortex part described by the singularities at $\Sigma $ such as 
to render all (closed) Wilson loops taken in the configuration 
$V\partial V^{\dagger } $ trivial. This means that $V\partial V^{\dagger } $ 
can be decomposed as\footnote{The minus sign in front of 
$a(\partial \Sigma )$ is a convention introduced for later notational
transparency.}
\begin{equation}
V(\Sigma ) \partial V^{\dagger } (\Sigma ) =
{\cal A} (\Sigma ) - a (\partial \Sigma )
\end{equation}
with the ideal vortex configuration ${\cal A} (\Sigma )$ localized on
a hypersurface $\Sigma $, and $a$ describing {\em thin vortices} localized 
on $\partial \Sigma $. The thin vortex field $a$ can be defined on 
$\Sigma $ as
\begin{eqnarray}
a|_{\Sigma } &=& - \lim_{\epsilon \rightarrow 0}
V(\Sigma , x_{\perp } , t=-\epsilon ) \partial
V^{\dagger } (\Sigma , x_{\perp } , t=-\epsilon ) \nonumber \\
&=& - \lim_{\epsilon \rightarrow 0}
V(\Sigma , x_{\perp } , t=\epsilon ) \partial
V^{\dagger } (\Sigma , x_{\perp } , t=\epsilon ) \ ,
\label{aonsig}
\end{eqnarray}
cf. eqs. (\ref{vdef1}). Away from $\Sigma $,
\begin{equation}
a=-V\partial V^{\dagger }
\label{aoffsig}
\end{equation}
since the ideal vortex configuration ${\cal A} (\Sigma )$ only has support 
on $\Sigma $.
\medskip

Before applying these singular gauge transformations to implement the
maximal center gauge in continuum Yang-Mills theory, a few further
properties should be noted for convenience. Given a general thin
vortex configuration $a$, a corresponding gauge transformation $V$
generating it can always be recovered via
\begin{equation}
V(y) = \exp \left( \int_{P(y_0 ,y)} a_{\mu } dx_{\mu } \right)
\end{equation}
using an arbitrary fixed starting point $y_0 $. Path ordering is
unnecessary, since $a$ was constructed to be an Abelian field.
Besides the trivial ambiguity in the choice of $y_0 $, which merely
leaves a constant phase undetermined in $V$, there is a more
important ambiguity, namely how the path $P(y_0 ,y)$ circumvents the
locations of the thin vortices contained in $a$. This choice precisely
determines the hypersurfaces $\Sigma $ at which $V$ jumps and which 
therefore describe the associated ideal vortex configuration
${\cal A} $.
\medskip

From this, it also becomes clear that the thin vortex configuration $a$
can be chosen independently of the precise singular surfaces $\Sigma $ 
describing the associated ideal vortex configuration. Merely the
boundary $\partial \Sigma $ must coincide with the location of the thin 
vortices contained in $a$. Indeed, given a transformation 
$V_1 (\Sigma_{1} )$ inducing a thin vortex field $a$, one can explicitly 
give another transformation $V_2 (\Sigma_{2} )$, with 
$\partial \Sigma_{1} = \partial \Sigma_{2} $, generating the same field 
$a$ as follows. Let $M$ denote the interior of the closed 
hypersurface\footnote{The notation ``$-\Sigma $'' is introduced to 
specify the orientation of a hypersurface. The hypersurface $-\Sigma $ 
is located on the same set of space-time points as the hypersurface 
$\Sigma $, but is associated with the reverse orientation.}
$\Sigma_{1} \cup -\Sigma_{2} $. In $D$ space-time dimensions, $M$ is 
$D$-dimensional. Then define
\begin{equation}
V_2 (k,\Sigma_{2} ) = \left\{
\begin{array}{rr}
Z(k) V_1 (k,\Sigma_{1} ) & \mbox{on } \, M \\
V_1 (k,\Sigma_{1} ) & \mbox{outside } \, M
\end{array}
\right.
\label{switch}
\end{equation}
This transformation generates the same thin vortex field $a$ as
$V_1 (\Sigma_{1} )$ and an ideal vortex configuration described by
the hypersurface $\Sigma_{2} $. In particular, using $V_1 (\Sigma_{1} )$
and $V_2 (\Sigma_{2} )$, one can deform the ideal vortex part of an
arbitrary gauge configuration without changing the configuration in
any other respect:
\begin{equation}
(A+ {\cal A} (\Sigma_{1} ) )^{V_1^{-1} V_2 } = A+ {\cal A} (\Sigma_{2} )
\label{noresch}
\end{equation}
This is the continuum analogue of a transformation from the center of 
the gauge group. It exclusively deforms ideal vortices.
Furthermore, the form of (\ref{switch}) shows that a homogeneous
rotation of any gauge field is independent of the location of the
singular hypersurfaces $\Sigma $,
\begin{equation}
V_1 (\Sigma_{1} ) A V_1^{\dagger } (\Sigma_{1} ) =
V_2 (\Sigma_{2} ) A V_2^{\dagger } (\Sigma_{2} )
\label{difsig}
\end{equation}
as long as $V_1 $ and $V_2 $ generate the same thin vortex field $a$, i.e.
in particular $\partial \Sigma_{1} =\partial \Sigma_{2} $.
\medskip

As a simple example, consider the $SU(2)$ transformation
\begin{equation}
V = \exp (\varphi T_3 )
\end{equation}
in the plane described by the polar coordinates $r,\varphi $ with
$\varphi \in [0,2\pi [ $. At $\varphi =0$, one has
\begin{equation}
V(r,\varphi =\epsilon ) V^{\dagger } (r, \varphi =2\pi -\epsilon ) 
= -1 +O(\epsilon )
\end{equation}
and, away from $\varphi =0$,
\begin{eqnarray}
a_{\varphi } &=& -\frac{1}{r} V\partial_{\varphi } V^{\dagger }
= \frac{1}{r} T_3 \\
a_r &=& -V\partial_{r} V^{\dagger } = 0
\end{eqnarray}
i.e. one observes an ideal vortex at $\varphi =0$ and a thin vortex
(cf. eq. (\ref{thinv})) at the origin, $r=0 $. By simply changing
the interval on which $\varphi $ is defined (in general, as a function
of $r$), one can arbitrarily deform the ideal vortex ${\cal A} (\Sigma )$
without changing $a(\partial \Sigma )$.
\medskip

\subsection{Continuum representation of center vortices}
In the continuum limit, ideal vortex configurations 
${\cal A}_{\mu } (k,\Sigma ,x)$ are specified by hypersurfaces 
$\Sigma $ (strings, sheets, or volumes in $D=2,3$ and $4$, respectively) 
which, when intersected by a Wilson line $C$, contribute a center 
element $Z(k)$ to the latter.
\medskip

An explicit (singular) gauge field representation of an ideal
vortex configuration in $D$ space-time dimensions is given by
\begin{equation}
{\cal A}_{\mu}(x,\Sigma ) = E\int_{\Sigma } d^{D-1} \tilde{\sigma}_{\mu}
\delta^{D} (x-\bar{x}(\sigma)) \, .
\label{idvorc}
\end{equation}
Here, $\Sigma $ describes the infinitely thin continuum limit of the
string, surface, or volume of nontrivial center element links making
up the ideal vortex configuration. $\Sigma $ is $D-1$-dimensional and
its boundary $S= \partial \Sigma $ gives the location of the thin vortex
$a_{\mu } (\partial \Sigma ,x)$ to which the ideal vortex configuration 
${\cal A}_{\mu } (\Sigma ,x)$ is gauge equivalent (see below). 
Furthermore, $\bar{x}_{\mu } (\sigma ) \equiv
\bar{x}_{\mu } (\sigma_{1} , \sigma_{2} ,\ldots ,\sigma_{D-1} )$ denotes
a parametrization of the $(D-1)$-dimensional hypersurface $\Sigma $.
Also,
\begin{equation}
d^{D-1} \tilde{\sigma }_{\mu } = \frac{1}{(D-1)! }
\epsilon_{\mu \alpha_{1} \ldots \alpha_{D-1} }
d^{D-1} \sigma_{\alpha_{1} \ldots \alpha_{D-1} }
\end{equation}
is the dual of the $(D-1)$-dimensional volume element
\begin{equation}
d^{D-1} \sigma_{\alpha_{1} \ldots \alpha_{D-1} } =
\epsilon_{k_{1} \ldots k_{D-1} }
\frac{\partial \bar{x}_{\alpha_{1} } }{\partial \sigma_{k_1 } }
\cdots \frac{\partial \bar{x}_{\alpha_{D-1} } }{\partial \sigma_{k_{D-1} } }
d\sigma_{1} \ldots d\sigma_{D-1} \ .
\end{equation}
Note that these volume elements also define an orientation of the vortex.
Like in the special case (\ref{thinv}), in general two different
orientations of vortices can be distinguished.
\medskip

On the other hand, the color structure of the vortex is encoded in $E$.
Note that the following discussion of the color factors $E$ applies equally 
to thin vortices, to be treated in detail further below. The different kinds
of vortex configurations introduced in this work are distinguished from 
one another only by their space-time structure, but they carry identical
color structure. It should furthermore be emphasized that the orientation 
of vortex flux is encoded already in the space-time structure, as mentioned 
above; this entails consequences regarding the set of color factors $E$
necessary to generate all possible distinct vortex fluxes. For example, 
the substitution of $E$ by $-E$ should not be considered to lead to
a new class of vortices; the configurations generated by this substitution
are already accounted for by keeping the original color structure $E$,
but reversing the orientation of the hypersurface $\Sigma $ in 
(\ref{idvorc}). As specified below, $E$ should be restricted to a certain 
fundamental domain.
\medskip

For the gauge group $SU(N)$, ideal center vortices in the continuum are 
given by eq. (\ref{idvorc}) with $E(k)=E_c (k) T_c $ satisfying
\begin{equation}
e^{-E(k)} = Z(k) \ , \ \ \ \ k=1,\ldots ,N-1
\label{roovdef}
\end{equation}
where only generators $T_c $ from the Cartan subalgebra are associated
with nonvanishing coefficients $E_c (k)$, and $Z(k)$ denotes the $N-1$
nontrivial center elements of the $SU(N)$ group. The $E_c (k)$ can be
considered as the components of vectors $E(k)$ in the Cartan subalgebra
which, up to a factor $2\pi $, coincide with the so-called co-weights
$\mu (k)$, i.e. $E(k)=2\pi \mu (k)$. The co-weights $\mu (k)$ are
dual to the simple roots and define the corners of the
fundamental domain of the $SU(N)$ algebra. They are defined in the
adjoint representation $(\hat{T}_{a} )_{bc} = f_{abc} $ by
\begin{equation}
e^{-\hat{E} } =1 \ , \ \ \ \hat{E} = E_c \hat{T}_{c}
\end{equation}
Here, mainly the gauge group $SU(2)$ will be considered, for which only
one nontrivial center element $Z=-1$ exists, associated with the color
structure
\begin{equation}
E=E_3 T_3 \ , \ \ \ E_3 =2\pi \ .
\label{su2e}
\end{equation}
For the gauge group $SU(3)$, there are two nontrivial center elements
\begin{equation}
Z(1) = e^{i2\pi /3} \ , \ \ \ \ \ Z(2) = e^{-i2\pi /3}
\end{equation}
which are generated from (\ref{roovdef}) by the vectors $E(1)$ and $E(2)$, 
whose non-vanishing components explicitly read
\begin{equation}
E_8 (1) = \frac{4\pi }{\sqrt{3} } \ , \ \ \ \ \
E_3 (2) = 2\pi \ , \ \ \ \ \
E_8 (2) = \frac{2\pi }{\sqrt{3} } \ .
\label{su3e}
\end{equation}
Note that both vectors $E(1)$ and $E(2)$ satisfy the normalization
\begin{equation}
\mbox{tr} (E(k))^2 = -\frac{1}{2} E_c (k) E_c (k) = -\frac{8\pi^{2} }{3}
\label{scprd}
\end{equation}
while
\begin{equation}
\mbox{tr} E(1) E(2) = -\frac{4\pi^{2} }{3} \ .
\label{scpro}
\end{equation}

With (\ref{roovdef}), one has for the Wilson line along a path $C$ in
the vortex configuration (\ref{idvorc})
\begin{equation}
W[ {\cal A} ] (C) = Z(k)^{I(C,\Sigma ) }
\label{wilsid}
\end{equation}
where
\begin{equation}
I(C,\Sigma ) = \int_{C} dx_{\mu } \int_{\Sigma }
d^{D-1} \tilde{\sigma }_{\mu } \delta^{D} (x-\bar{x} (\sigma ) )
\end{equation}
is the intersection number between $C$ and $\Sigma $ in $D$ dimensions.
For closed paths $C$, i.e. for Wilson loops, this equals the linking
number between $C$ and $S=\partial \Sigma $,
\begin{equation}
I(C,\Sigma ) = L (C,S=\partial \Sigma ) \ .
\end{equation}
In particular, ${\cal A} $ therefore satisfies the properties
characterizing an ideal vortex configuration as formulated at the
beginning of this section. As already mentioned in connection with
eq. (\ref{noresch}), the sheet $\Sigma $ can be arbitrarily deformed by 
a (singular) Abelian gauge transformation, while keeping its boundary
$S=\partial \Sigma $ fixed; this is entirely analogous to the lattice, 
where the stack of links $U = -1$ (in the case of $SU(2)$)
forming the hypersurface $\Sigma $ attached to the center vortex
$S = \partial\Sigma$ can be deformed by center gauge transformations.
A more explicit construction of this transformation is given further
below in eq. (\ref{swatch}).
\medskip

On the other hand, in order to find a general representation of thin
vortex configurations, consider the following gauge transformation,
\begin{equation}
V(k,\Sigma , x) = \exp \left( E(k) \Omega (\Sigma ,x) \right)
\label{vvonsol}
\end{equation}
where
\begin{equation}
\Omega (\Sigma ,x) = \frac{1}{\omega_{D-1} }
\int_{\Sigma } d^{D-1} \tilde{\sigma }_{\mu }
\frac{\bar{x}_{\mu } (\sigma ) -x_{\mu } }{(\bar{x} (\sigma ) -x)^D }
\label{solang}
\end{equation}
is the solid angle taken up by the hypersurface $\Sigma $ when viewed
from $x$. Here,
\begin{equation}
\omega_{D-1} = \frac{2\pi^{D/2} }{\Gamma (D/2)}
\end{equation}
is the area of the unit sphere $S_{D-1} $ in $D$ dimensions, so that the
solid angle as defined above is normalized to unity for a point $x$
inside a closed surface $\Sigma $. Note that the solid
angle is defined with a sign depending on the orientation of $\Sigma $
as rays emanating from $x$ pierce $\Sigma $. A deformation of $\Sigma $
keeping its boundary $\partial \Sigma $ fixed leaves $\Omega (\Sigma ,x)$
invariant unless $x$ crosses $\Sigma $. In the latter case,
$\Omega (\Sigma ,x)$ changes by an integer.
\medskip

For later convenience, note also that the solid angle (\ref{solang})
can be represented as
\begin{equation}
\Omega (\Sigma ,x) = \int_{\Sigma } d^{D-1} \tilde{\sigma }_{\mu }
\partial_{\mu }^{x} D (x-\bar{x} (\sigma ))
\end{equation}
where $D(x)$ is the Green's function of the $D$-dimensional Laplacian,
\begin{equation}
-\partial_{\mu }^{x} \partial_{\mu }^{x} D(x-x^{\prime } )
=\delta^{4} (x-x^{\prime } ) \ .
\label{ddlap}
\end{equation}
For a closed surface $\Sigma =\partial M$ the solid angle agrees
with the characteristic function $\chi (M,x)$ of $M$ by Gau\ss ' theorem,
\begin{equation}
\Omega (\Sigma ,x) = \int_{M} d^D \sigma \delta^{D} (x-\bar{x} (\sigma ) )
=\chi (M,x)
\end{equation}
i.e. any point inside $\Sigma =\partial M$ is associated with a solid 
angle $1$, any point outside $\Sigma $ with a vanishing solid angle,
\begin{equation}
\chi (M,x) = \left\{
\begin{array}{rr}
1 & x\in M \\
0 & \mbox{otherwise}
\end{array}
\right.
\label{chidef}
\end{equation}
In this respect, the normalized solid angle may also be interpreted as the
linking number of the point $x$ and the hypersurface $\Sigma $.
\medskip

Consider now the gauge field induced by the gauge transformation 
(\ref{vvonsol}) defined by the solid angle $\Omega (\Sigma ,x)$,
\begin{equation}
V(k,\Sigma ,x) \partial_{\mu } V^{\dagger } (k,\Sigma ,x) =
-E(k) \partial_{\mu } \Omega (\Sigma ,x) \ .
\label{vbsp}
\end{equation}
Obviously, one has
\begin{equation}
\oint_{C} dx_{\mu } \, \partial_{\mu } \Omega (\Sigma ,x) =0
\end{equation} 
for any closed curve $C$, so that any Wilson loop receives the trivial
contribution
\begin{equation}
W [V\partial V^{\dagger } ] (C) =1 \ .
\end{equation}
Now one can show that indeed
\begin{equation}
V(k,\Sigma ,x) \partial_{\mu } V^{\dagger } (k,\Sigma ,x) =
{\cal A}_{\mu } (k,\Sigma ,x) - a_{\mu } (k,\partial \Sigma ,x)
\label{tvex}
\end{equation}
where $a_{\mu } $ and ${\cal A}_{\mu } $ are the thin and the ideal 
vortex fields, respectively, describing the same magnetic flux localized 
on $\partial \Sigma $. While ${\cal A}_{\mu } $ has already been defined
by eq. (\ref{idvorc}), an explicit representation for the thin vortex 
$a_{\mu } (k,\partial \Sigma ,x) $ in (\ref{tvex}) will be derived below.
For this purpose, consider
\begin{equation}
\partial_{\mu } \Omega (\Sigma ,x) =
\int_{\Sigma } d^3 \tilde{\sigma }_{\nu }
\partial_{\mu } \partial_{\nu } D(x-\bar{x} (\sigma ) )
\label{proo}
\end{equation}
and write
\begin{equation}
\partial_{\mu } \partial_{\nu } = \delta_{\mu \nu } \partial^{2}
- (\delta_{\mu \nu } \partial^{2} - \partial_{\mu } \partial_{\nu } )
\end{equation}
i.e. express the longitudinal projector $L_{\mu \nu } =
\partial_{\mu } \partial_{\nu } /\partial^{2} $ as
$\delta_{\mu \nu } - {\cal P}_{\mu \nu } $, where
${\cal P}_{\mu \nu } = \delta_{\mu \nu } -
\partial_{\mu } \partial_{\nu } /\partial^{2} $
is the transversal projector. Inserting this relation into eq. (\ref{proo})
and using the definition of the Green's function (\ref{ddlap}), one obtains
\begin{equation}
\partial_{\mu } \Omega (\Sigma ,x) =
-\int_{\Sigma } d^{D-1} \tilde{\sigma }_{\mu }
\delta^{D} (x-\bar{x} (\sigma ) ) -
\int_{\Sigma } d^{D-1} \tilde{\sigma }_{\nu }
(\delta_{\mu \nu } \partial^{2} - \partial_{\mu } \partial_{\nu } )
D(x-\bar{x} (\sigma ) ) \ .
\end{equation} 
Upon inserting this relation into (\ref{vbsp}), the first term produces
precisely the ideal vortex field ${\cal A} (k,\Sigma ,x)$ as defined
in eq. (\ref{idvorc}). To show that the second term yields the thin vortex 
field, it is necessary to express the transversal differential operator as
\begin{equation}
\delta_{\mu \nu } \partial^{2} - \partial_{\mu } \partial_{\nu }
=\frac{1}{2} \epsilon_{\mu \kappa \alpha \beta }
\epsilon_{\nu \lambda \alpha \beta }
\partial_{\kappa } \partial_{\lambda } \ .
\end{equation}
After straightforward calculation, using Stokes' theorem, one eventually
obtains
\begin{equation}
-\int_{\Sigma } d^{D-1} \tilde{\sigma }_{\nu }
(\delta_{\mu \nu } \partial^{2} - \partial_{\mu } \partial_{\nu } )
D(x-\bar{x} (\sigma ) ) =
-\int_{\partial \Sigma } d^{D-2} \tilde{\sigma }_{\mu \kappa }
\partial_{\kappa }^{\bar{x} } D(x-\bar{x} (\sigma ) )
\label{projaus}
\end{equation}
Inserting this into eq. (\ref{vbsp}), one finds the thin vortex for
arbitrary $\partial \Sigma $,
\begin{equation}
a_{\mu } (k,\partial \Sigma ,x) = -E(k)
\int_{\partial \Sigma } d^{D-2} \tilde{\sigma }_{\mu \kappa }
\partial_{\kappa }^{\bar{x} } D(x-\bar{x} (\sigma ) )
\label{arbthv}
\end{equation}
In Appendix \ref{appthv}, it is shown explicitly that 
$a_{\mu } (k,\partial \Sigma , x)$ in eq. (\ref{arbthv}) indeed represents 
a center vortex, namely
\begin{equation}
\oint_{C} dx^{\mu } a_{\mu } (k,\partial \Sigma , x) =
E(k) L(C,\partial \Sigma )
\end{equation}
where $L(C,\partial \Sigma )$ is the linking number between $C$ and 
$\partial \Sigma $.
\medskip

Note that, unlike the ideal vortex ${\cal A} (\Sigma ,x)$, the thin
vortex $a(\partial \Sigma ,x)$ depends only on the boundary
$\partial \Sigma $, where the magnetic flux associated with the vortex
is located.
\medskip

It is now easy to see that a gauge transformation which deforms
the singular sheet $\Sigma_{1} $, describing an ideal vortex, into
another sheet $\Sigma_{2} $ while keeping the boundary $\partial \Sigma $ 
fixed is given by
\begin{eqnarray}
V(k,\Sigma_{1} ,\Sigma_{2} ,x) &=& V^{-1} (k,\Sigma_{1} ,x) 
V (k,\Sigma_{2} ,x) \\
&=& \exp \left[ -E(k) (\Omega (\Sigma_{1} ,x)-\Omega (\Sigma_{2} ,x)
) \right]
\label{swatch}
\end{eqnarray}
Since the boundaries of $\Sigma_{1} $ and $\Sigma_{2} $ match, 
$\partial \Sigma_{1} = \partial \Sigma_{2} $, the thin vortices
arising from $V(\Sigma_{1} ) \partial V^{\dagger } (\Sigma_{1} )$
and $V(\Sigma_{2} ) \partial V^{\dagger } (\Sigma_{2} )$ cancel and one 
obtains
\begin{equation}
( {\cal A} (k,\Sigma_{1} ,x) )^{V(k,\Sigma_{1} , \Sigma_{2} ,x)}
={\cal A} (k,\Sigma_{2} ,x) \ .
\end{equation}
One can also easily convince oneself that the transformation (\ref{swatch})
agrees with the one introduced in section \ref{singgt}, eq. (\ref{switch}).
Since $\Sigma_{1} \cup (-\Sigma_{2} ) = \partial M$ represents the closed
orientable surface of some volume $M$, application of Gau\ss ' theorem
yields
\begin{eqnarray}
\Omega (\Sigma_{1} ,x)-\Omega (\Sigma_{2} ,x) &=&
-\int_{\Sigma_{1} } d^{D-1} \tilde{\sigma }_{\mu } 
\partial_{\mu }^{\bar{x} } D(x-\bar{x} (\sigma ) ) \\
& & \ \ \ \ \ \ \ \ \ \ \ \ \ \ \ \ \ \
+ \int_{\Sigma_{2} } d^{D-1} \tilde{\sigma }_{\mu }
\partial_{\mu }^{\bar{x} } D(x-\bar{x} (\sigma ) ) \nonumber \\
&=& -\int_{\Sigma_{1} \cup (-\Sigma_{2} ) } d^{D-1} \tilde{\sigma }_{\mu }
\partial_{\mu }^{\bar{x} } D(x-\bar{x} (\sigma ) ) \\
&=& \int_{M} d^D \tilde{\sigma } 
(-\partial_{\mu }^{\bar{x} } \partial_{\mu }^{\bar{x} } )
D(x-\bar{x} (\sigma ) ) \ .
\end{eqnarray}
Using the definition of the Green's function of the $D$-dimensional
Laplacian (\ref{ddlap}), one obtains
\begin{equation}
\Omega (\Sigma_{1} ,x)-\Omega (\Sigma_{2} ,x)
= \int_{M} d^D \tilde{\sigma } \delta^{D} (x-\bar{x} (\sigma ) )
= \chi (M,x)
\end{equation}
where $\chi (M,x)$ is the characteristic function of $M$ defined in 
eq. (\ref{chidef}), and $\partial M=\Sigma_{1} \cup (-\Sigma_{2} )$.
Hence, the gauge transformation (\ref{swatch}) can be expressed as
\begin{eqnarray}
V(k,\Sigma_{1} ,\Sigma_{2} ,x) &=&
\exp \left[ -E(k) \chi (M,x) \right] \\
&=& Z(k)^{\chi (M,x) } \ .
\end{eqnarray}
Thus, one recovers precisely the transformation already introduced 
in (\ref{switch}).
\medskip

\subsection{Continuum limit of the Maximal Center gauge}
\label{congsec}
On the basis of the groundwork laid out in the previous sections,
it is now straightforward to construct the analogue of the maximal 
center gauge for continuum Yang-Mills fields. The transformations 
$V(\Sigma )$ characterized by (\ref{vdef1}) allow to carry out precisely
step 1.) of the maximal center gauge fixing procedure, as specified
in section \ref{sepsec}, in the continuum. Given an initial smooth
gauge field $A$, the transformed field $A^{V (\Sigma )} $ contains ideal 
vortices described by hypersurfaces $\Sigma $ of one's choice.
Furthermore, in analogy to the lattice, where taking the coset part
of a configuration amounts to leaving away all ideal vortices
${\cal A} (k,\Sigma ,x)$, one can define the stripped configuration
\begin{equation}
(A^V )^{\prime } = A^V - {\cal A} (k,\Sigma ) = VAV^{\dagger } 
- a (k,\partial \Sigma )
\label{stripa}
\end{equation}
with the thin vortex field $a$ defined in terms of $V$ as in
(\ref{aonsig}) and (\ref{aoffsig}), an explicit representation being
given by eq. (\ref{arbthv}). This stripped configuration
induces the same behavior for arbitrary Wilson line integrals as
the configuration $A^V $, except for the center element jumps at
the hypersurfaces $\Sigma $. In fact, $(A^V )^{\prime } $ depends
only on the thin vortex configuration $a$, with thin vortices
at $\partial \Sigma $, and not on the specific choice of the entire
hypersurface $\Sigma $, cf. eq. (\ref{difsig}). It is therefore the 
precise analogue of the coset part in the lattice case. Note that 
$(A^V )^{\prime } $ is still singular on $\partial \Sigma $.
\medskip

According to step 2.) of the maximal center gauge fixing procedure as
specified in section \ref{sepsec}, one must now consider all coset
gauge transformations of $(A^V )^{\prime } $ such as to maximize the
gauge fixing functional. As explained in the previous section, this step
can be straightforwardly translated to the continuum, where gauge
fields are subject to arbitrary local coset transformations.
Expressing therefore the coset link configuration in the continuum 
limit, cf. eq. (\ref{cosexp}),
\begin{equation}
U^{\prime }_{\mu  } =
1 - l A_{\mu }^{\prime } +\frac{l^2 }{2} (A_{\mu }^{\prime } )^{2} - \ldots
\end{equation}
the maximal center gauge fixing condition (\ref{sepgf}) can be rewritten as
\begin{equation}
\max_{\Sigma } \max_{G} \int d^D x \, g \left(
\mbox{tr} \left( (A^{V(\Sigma ) }_{\mu } )^{\prime \, G}
(A^{V(\Sigma ) }_{\mu } )^{\prime \, G} \right) \right)
\label{cgcont}
\end{equation}
where the arbitrary gauge fixing function has also been redefined,
\begin{equation}
g(t) = \frac{1}{l^D } \bar{g} \left( 2+\frac{l^2 }{2} t \right) \ .
\label{redefg}
\end{equation}
Like $\bar{g} $, $g$ should be a monotonously rising function. Note
that in (\ref{cgcont}), the maximization is, as written, over all
hypersurfaces $\Sigma $, and not over all possible associated
transformations $V(\Sigma )$. For every $\Sigma $, there are many
possible $V(\Sigma )$. Only one $V(\Sigma )$ must be constructed and
inserted in (\ref{cgcont}) for every choice of $\Sigma $.
\medskip

Inserting (\ref{stripa}), the gauge fixing condition can be written as
\begin{equation}
\max_{\Sigma } \max_{G} \int d^D x \, g \left(
\mbox{tr} \left( GV(A-a)V^{\dagger } G^{\dagger } + G\partial G^{\dagger }
\right)^{2} \right) \ .
\label{mcgcalt}
\end{equation}
This expression can be further simplified such as to eliminate all
explicit reference to the singular transformations $V$. In the following
calculation, there is a slight subtlety. Namely, the integrand in
(\ref{mcgcalt}) is continuous at hypersurfaces $\Sigma $; it was
precisely constructed to be so. Thus, one may replace its value
on hypersurfaces $\Sigma $ by the limit as one approaches $\Sigma $.
In the following, the integrand in (\ref{mcgcalt}) will therefore be
evaluated outside of $\Sigma $ and its value on $\Sigma $ is then
defined by continuity. In particular, this means that one may set
$V\partial V^{\dagger } =-a$, cf. eq. (\ref{aoffsig}).
\medskip

By writing $G=V\bar{G} V^{\dagger } $, the argument of the gauge function
$g$ in (\ref{mcgcalt}) becomes
\begin{eqnarray}
& & \hspace{-1cm} 
\mbox{tr} \left( V\bar{G} (A-a) \bar{G}^{\dagger } V^{\dagger } 
+V\bar{G} (V^{\dagger } \partial V) \bar{G}^{\dagger } V^{\dagger }
+V(\bar{G} \partial \bar{G}^{\dagger } ) V^{\dagger }
+V\partial V^{\dagger } \right)^{2} \\
& & \ \ \ = \mbox{tr} \left( V\bar{G} A \bar{G}^{\dagger } V^{\dagger }
+V(\bar{G} \partial \bar{G}^{\dagger } ) V^{\dagger }
+V\partial V^{\dagger } \right)^{2} \\
& & \ \ \ = \mbox{tr} \left( V A^{\bar{G} } V^{\dagger } - a \right)^{2} \\
& & \ \ \ = \mbox{tr} \left( A^{\bar{G} } - a \right)^{2}
\end{eqnarray}
where the fact that $V$ and $a$ commute was used. Note that $\bar{G} $ is
continuous on $\Sigma $, since it is quadratic in $V$ and hence
insensitive to the center phase $V$ picks up at hypersurfaces $\Sigma $.
Thus, it can still be considered a coset gauge transformation. Vice versa,
any coset transformation $\bar{G} $ is associated with a coset
transformation $G$. Therefore, the gauge fixing condition (\ref{mcgcalt})
can be rewritten as
\begin{equation}
\max_{\Sigma } \max_{\bar{G} } \int d^D x \, g \left(
\mbox{tr} \left( A^{\bar{G} } - a
\right)^{2} \right)
\label{mcgc}
\end{equation}
with the coset maximization running over all coset transformations
$\bar{G} $.
\medskip

In this form, maximal center gauge fixing is revealed more clearly as an 
approximation problem in the mathematical sense; it corresponds to 
approximating the gauge orbit of a given gauge configuration $A$ as well 
as possible by thin vortex configurations $a$. The precise metric defining 
how close a thin vortex field $a$ is to the gauge orbit of $A$ is
encoded in the arbitrary function $g$. Thus, the continuum formulation of 
the maximal center gauge provides a deeper understanding of the empirical 
fact drawn from lattice experiments \cite{deb97}, namely that center 
projection vortices provide a rough localization of thick vortex structures 
present in full lattice Yang-Mills field configurations. Note that, in 
mathematical terms, the space of thin vortex fields $a$ has no scalar 
product or orthogonality properties. Eq. (\ref{mcgc}) specifies a norm 
which serves to define an approximation problem within this space, 
namely maximal center gauge fixing.
\medskip

Note that the gauge fixing functional (\ref{mcgc}) only depends on the 
thin vortex configuration $a$, with vortices localized on 
$\partial \Sigma $, and not on the whole choice of $\Sigma $.
This is the continuum counterpart of the lattice 
observation that the maximal center gauge leaves the center of the 
gauge group unfixed, i.e. does not depend on the center part of a 
configuration.
\medskip

As a last step, it is interesting to recast the coset gauge fixing 
condition in differential form as follows. Consider $\Sigma $ and an 
associated thin vortex field $a(\partial \Sigma )$ as given. Then a gauge 
field $A$ satisfies the condition of providing the maximal value of the 
gauge fixing functional under the action of arbitrary coset 
transformations $\bar{G} $ if the value of the gauge fixing functional 
(\ref{mcgc}) is stationary under infinitesimal transformations 
$\bar{G} =\exp (-\theta ) \approx 1-\theta $,
\begin{eqnarray}
0 &=& \frac{\delta }{\delta \theta^{a} (y) } \int d^D x \, g \left(
\mbox{tr} \left( [A,\theta ] + \partial \theta -a \right)^{2} \right) \\
&=& \int d^D x \, g^{\prime } \left( \mbox{tr} (A-a)^2 \right)
(A^c -a^c ) (-\partial \delta^{ac} -A^d f^{dac} ) \delta^{D} (x-y) \\
&=& \partial_{\mu } \left[ g^{\prime } \left( \mbox{tr} (A-a)^2 \right)
(A_{\mu }^{a} -a_{\mu }^{a} ) \right] -
g^{\prime } \left( \mbox{tr} (A-a)^2 \right) f^{adc} A^d a^c
\end{eqnarray}
where the prime on $g$ denotes the derivative with respect to the argument.
By multiplying with generators $T_a $, this reduces to
\begin{eqnarray}
0 = F[A,a]
&=& 2\, g^{\prime \prime } \left( \mbox{tr} \left( A-a \right)^{2} \right)
\cdot \mbox{ tr} \left( (A_{\nu } - a_{\nu} ) \partial_{\mu  }
(A_{\nu } - a_{\nu} ) \right) \cdot (A_{\mu } - a_{\mu} ) \nonumber \\ 
& & + g^{\prime } \left( \mbox{tr} \left( A-a \right)^{2} \right)
\left( [ \partial_{\mu } + a_{\mu }, A_{\mu } ] - \partial_{\mu } a_{\mu }
\right) 
\label{gaucon}
\end{eqnarray}
In the simple case $g(t)=t$, the gauge condition 
simplifies to 
\begin{equation}
[ \partial_{\mu } + a_{\mu }, A_{\mu } ] - \partial_{\mu } a_{\mu } =0
\end{equation}
If one in particular chooses the thin vortex configuration to satisfy
the Landau gauge, $\partial_{\mu } a_{\mu } =0$, then this condition
is nothing but the background gauge with the background given by
the thin vortex field $a$. Note that the thin vortex configuration 
(\ref{arbthv}) satisfies the Landau gauge, as can be inferred from
eq. (\ref{projaus}).
\medskip

Furthermore, if one has picked out of the gauge orbit of a given
configuration $A$ the element $A[a]$ which satisfies (\ref{gaucon}) for
fixed arbitrary $a$ (coset gauge fixing), then the remaining part of 
the maximal center gauge fixing procedure, i.e. finding the optimal 
$a$, reduces to solving
\begin{equation}
\max_{\Sigma } \int d^D x \, g \left( \mbox{tr} (A[a] -a)^2 \right) \ .
\label{einfgc}
\end{equation}
As mentioned repeatedly above, all choices of $\Sigma $ in (\ref{einfgc}) 
corresponding to the same $\partial \Sigma $ are degenerate due to the 
unfixed center part of the $SU(2)$ gauge freedom. Also, for every 
$\partial \Sigma $, it is only necessary to construct one particular 
$a(\partial \Sigma )$, cf. the discussion after eq. (\ref{redefg}).
\medskip

Finally, having obtained an optimal hypersurface $\Sigma $, {\em center
projection} simply means replacing the full gauge field by the
corresponding ideal vortex configuration ${\cal A} (\Sigma )$,
cf. eq. (\ref{idvorc}). This center projected configuration can then
e.g. be used to evaluate observables such as the Wilson loop, cf.
eq. (\ref{wilsid}).
\medskip

\subsection{Remarks on the center gauges in continuum Yang-Mills theory}
\subsubsection{Revisiting the thick vortex}
\label{revthsec}
In general, the gauge fixing condition (\ref{gaucon}) is not solvable
by analytical means; at this stage, the value of the continuum formulation
of the maximal center gauge lies more on a conceptual than on a practical
level. However, it is worthwhile to briefly reexamine some properties of the
examples presented in section \ref{examse} in the lattice framework. Namely,
consider approximating (the gauge orbit of) a given thick $SU(2)$ vortex
configuration centered on the origin of a two-dimensional space-time
plane, cf. eq. (\ref{thickv}),
\begin{equation}
A = T_3 \frac{f(\sqrt{x^2 +y^2 } )}{x^2 +y^2 } 
(y {\bf e}_x -x {\bf e}_y ) \ , \ \ \ \ \ \ \partial A =0
\end{equation}
by a thin vortex centered at the point $x_0 {\bf e}_x $,
\begin{equation}
a = T_3 \frac{1}{(x-x_0 )^2 +y^2 } (y {\bf e}_x -(x-x_0 ) {\bf e}_y )
\end{equation}
where the offset $x_0 $ is still to be varied; 
$ {\bf e}_x , {\bf e}_y $ denote the Cartesian unit vectors in the 
plane. For definiteness, the following treatment was carried 
out using a profile function $f(r)=\theta (r-R)$, i.e. a ring-shaped 
thick vortex of radius $R$ similar to the one considered in 
section \ref{examse}. In the case $x_0 =0$, the gauge condition 
(\ref{gaucon}) is satisfied by these configurations, for 
arbitrary gauge fixing function $g$. For $x_0 \neq 0$, the 
condition (\ref{gaucon}) is in general not satisfied. 
Nevertheless, in terms of the variational form (\ref{mcgc}) 
of the gauge fixing condition, one has the relation
\begin{equation}
\int d^D x \, g \left( \mbox{tr} (A-a)^2 \right) \leq \max_{\bar{G} } 
\int d^D x \, g \left( \mbox{tr} (A^{\bar{G} } -a)^2 \right)
\label{coexab}
\end{equation}
where equality is guaranteed for $x_0 =0$. Now, it is possible to find 
gauge fixing functions $g$, e.g. $g=-\tanh (R^4 t^2 )$, for which the 
quantity
\begin{equation}
\int d^D x \, g \left( \mbox{tr} (A-a)^2 \right)
\end{equation}
appearing on the left hand side of (\ref{coexab}), viewed as a function
of $x_0 $, exhibits a minimum at $x_0 =0$. In view of (\ref{coexab}),
this implies that also the quantity on the right hand side,
\begin{equation}
\max_{\bar{G} } \int d^D x \, g \left( \mbox{tr} (A^{\bar{G} } -a)^2
\right)
\end{equation}
exhibits a minimum at $x_0 =0$ (remember that equality is guaranteed in 
(\ref{coexab}) for $x_0 =0$). Therefore, according to the variational
form of the gauge fixing condition (\ref{mcgc}), the optimal location
of the approximating thin vortex configuration $a$, parametrized by
$x_0 $, is achieved for $x_0 \neq 0$. This confirms the phenomenon
observed in the lattice examples in section \ref{examse}, that the
thin vortex induced by applying the center gauge fixing procedure to a
radially symmetric thick vortex configuration does not necessarily
appear concentrically with the original thick vortex. This property 
thus does not constitute a lattice artefact, but persists also in the 
continuum theory.
\medskip

As a final remark, note that the simple choice of gauge fixing function 
$g(t)=t$ leads to a trivial optimal thin vortex configuration $a=0$ for 
any smooth initial configuration $A$. This is due to the fact that a thin 
vortex field diverges as $1/r$, where $r$ is the distance from the vortex, 
and therefore tr$(A-a)^2 $ diverges as $-1/r^2 $ at the positions of the
thin vortices. Therefore, in the case $g(t)=t$, the only choice of
thin vortex field which does not render the gauge fixing functional
(\ref{einfgc}) divergent is $a=0$. Consequently, $a=0$ trivially
turns out to be the optimal thin vortex approximation, for any
smooth $A_{\mu } (x) $. The choice
$g(t)=t$ for the gauge fixing function thus at first sight does not seem
very useful, since it cannot lead to a faithful representation of the
vortex content of the original smooth gauge field $A$. The same is true
for any $g(t)$ which diverges as fast as, or faster than, $g(t)=t$
as $t\rightarrow \infty $.
\medskip

\subsubsection{Remarks on the Direct Maximal Center Gauge}
\label{dmcgsec}
The choice of gauge fixing function $g(t)=t$ is the continuum analogue
of a variety of lattice gauge fixing functions $\bar{g} (x)$,
cf. eq. (\ref{redefg}) in the limit $l\rightarrow 0$. Among these
is the case $\bar{g} (x) = x^2 $ (after leaving away irrelevant constants).
This is the standard ``Direct Maximal Center Gauge'' 
\cite{deb97aug},\cite{giedt},\cite{montero}. Strictly speaking, therefore, 
this gauge does not have a useful continuum limit on the level of gauge 
fixing individual smooth continuum gauge field configurations. The question 
arises how this gauge leads to a finite center projection vortex density in 
the continuum limit of the lattice formulation, as evidenced by the 
renormalization group scaling behavior of this density \cite{kurt}
(note erratum in \cite{tempv}), see also \cite{giedt}.
Several possible explanations come to mind.
\medskip

For one, in a lattice calculation, one never takes the continuum limit
of the field configurations at an intermediate stage. Instead, one
operates with regularized lattice configurations and only extrapolates
observables at the end of the calculation to the continuum limit.
Thus, on the lattice, one always stays away a certain minimal distance
from the $1/r$-singularity of the continuum thin vortex, i.e. one
never probes the $t\rightarrow \infty $ behavior of $g(t)$. More
precisely, one cannot make a distinction between $g(t)=t$ and a
``capped'' version $g_T (t) = t\theta (T-t)+T\theta (t-T)$, with a
sufficiently large value of $T$. Of course, as the lattice spacing
is taken to zero, $T$ will have to diverge if $g$ and $g_T $ are to
remain equivalent. Now, it is entirely possible that the choice
$g(t)=t$ leads to scaling violations of the projection vortex density on 
very fine lattices, such that this density ultimately tends to zero.
On the other hand, a capped version $g_T (t)$ as above, with large but
fixed $T$, allows a finite projection vortex density for smooth 
continuum configurations. Possibly, present lattices are still too coarse 
to reveal the above scaling violations and distinguish between $g(t)=t$
and $g_T (t)$ with large fixed $T$. An observation which supports this
possibility is that on present lattices, one still finds in typical
configurations an abundance of negative plaquettes, i.e. structures
indistinguishable from thin vortices on the scale of the lattice spacing.
On the other hand, the appeal of this explanation is lessened by the
fact that it calls into question the relevance of the scaling behavior
observed for the projection vortex density; usually, such behavior is 
considered a firm indication that the continuum limit of an observable 
can be extrapolated with confidence.
\medskip

A second explanation can be based on the fact that functional integrals
in field theory are dominated, for entropy reasons, by configurations
with infinite action, not by smooth configurations \cite{rivers}.
For instance, if typical configurations $A$ contain thin vortices,
despite the fact that they are associated with a divergent Yang-Mills
action, then in the process of maximal center gauge fixing, i.e.
approximating the configurations by thin vortices, one would indeed
introduce a nontrivial thin vortex configuration $a$ to cancel all
thin vortices in the combination $A-a$. This would happen also for the
choice of gauge fixing function $g(t)=t$. However, it has been argued
that thin vortices are too singular to survive the continuum limit
of lattice Yang-Mills theory \cite{vortvan}. In this simple
form, this second explanation is therefore questionable.
\medskip

However, a third explanation which is a slightly more sophisticated
variant of the second one is possible. When fixing to the maximal
center gauge on the lattice, one never finds the exact maximum of
the gauge fixing functional, but only an approximation of this
maximum. Put another way, consider introducing the (negative)
gauge fixing functional as a weight into the Yang-Mills partition
function,
\begin{equation}
S_{GF} = -q \int d^D x \, g \left( \mbox{tr} (A[a] -a)^2 \right)
\end{equation}
cf. eq. (\ref{einfgc}). Then, in practice, one does not insist on the
limit $q\rightarrow \infty $, which would imply finding the exact
minimum of $S_{GF} $; instead one may be satisfied with a finite
value of $q$. In this case, a competition between the gauge fixing term
and the Yang-Mills action (and also a competition between different
Gribov copies, i.e. relative minima of $S_{GF} $) becomes possible. 
Even in a case such as $g(t)=t$, where the gauge fixing term 
in general will diverge for nontrivial thin vortex configurations 
$a$, these divergences may be swamped by the infinite action 
of typical Yang-Mills field configurations. Note that for 
this to work, only Yang-Mills configurations considerably 
less singular than the thin vortices invoked in the second 
explanation above are necessary. E.g., in the case of the 
gauge fixing function $g(t)=t$, the gauge fixing term for a
nontrivial thin vortex configuration $a$ will diverge as
$\int dr\, r (1/r^2)$, where $r$ denotes the distance from the thin
vortex. This is independent of the dimension of space-time; the
coordinates perpendicular to a vortex always span two dimensions.
On the other hand, in four space-time dimensions, already a gauge
field configuration which behaves as $1/r$ in the vicinity of a point
singularity will induce an action diverging as $\int dr\, r^3 (1/r^4)$;
a divergence of the same degree is induced by a $1/\sqrt{r} $ line
singularity or a $\ln (r)$ sheet singularity in four space-time 
dimensions.
\medskip

\subsubsection{Alternative vortex gauges}
\label{altvog}
The remarks above illustrate that the continuum maximal center gauge
in the form derived in section \ref{congsec} in a sense constitutes
excessive rigor. In a {\em dynamical} calculation, one is forced to
introduce a regularization of divergences in the ultraviolet; this
regularization in particular also smears out the thin vortex
configurations $a(\partial \Sigma )$, used in maximal center gauge
fixing, into thick vortices with a thickness related to the ultraviolet
cutoff. In view of this regularization, dictated by the dynamics,
it is not mandated to consider at an intermediate stage the behavior of
gauge field configurations as the cutoff is removed. Instead, the cutoff
must be kept finite throughout the calculation, and only the continuum
limit of physical observables, as encoded in their renormalization 
group behavior, is relevant.
\medskip

This suggests an alternative formulation of the gauge fixing procedure 
in which one approximates a given gauge configuration as well as possible
by thick vortices. Specifically, one can generalize the gauge
condition (\ref{gaucon}) and the remaining maximization problem 
(\ref{einfgc}) by allowing the approximating configurations
$a(\partial \Sigma )$ to comprise thick vortices. Of course, such
a generalization clouds the original idea of the maximal center gauge,
namely keeping the center part of the gauge freedom on the lattice
unfixed; instead, it operates purely with continuum, i.e. coset
configurations. Therefore, such gauges should be more appropriately 
called maximal vortex gauges rather than center gauges.
\medskip

In such vortex gauges, one has an additional freedom in the choice of the
thickness of the approximating vortex configurations $a(\partial \Sigma )$.
On the one hand, one may choose the thickness to be related to the
ultraviolet cutoff with which one regulates the theory; for example,
the thickness may simply be the lattice spacing. This form would be 
entirely equivalent to the maximal center gauge in the lattice theory, 
since, at the level of the regularized theory, there is no distinction 
between such a thick vortex and a truly thin one. On the other hand, one 
may choose the thickness to be a fixed physical quantity\footnote{Some
phenomenological implications of a physical vortex thickness were
discussed in \cite{adj1},\cite{adj2}.} (which can be
varied to optimally match the vortex profile preferred by the Yang-Mills
dynamics). In this form, the thick vortex profile can also take 
over the role of the gauge fixing function $g$. For example, as evidenced
in the examples given further above, the gauge fixing function $g$ can
act as a regulator for the divergences introduced into the gauge fixing
functional by the singularities of the thin vortices. In a vortex
gauge with thick vortices, this is already achieved by the vortex 
thickness. In many respects, the vortex profile in the vortex gauge has
a similar effect as the gauge fixing function $g$ in the maximal center
gauge.
\medskip

As an aside, note that in a consistently regularized theory, one should 
also use regularized versions of the ideal vortex configurations
${\cal A}$, cf. (\ref{idvorc}), which one simply obtains by using 
the ultraviolet cutoff of the theory to smooth out the $\delta $-functions 
in (\ref{idvorc}), cf. eq. (\ref{regidv}).
\medskip

Lastly, a certain aspect of continuum thin (or, equivalently, ideal)
vortices has not yet been discussed in detail, namely the possibility 
that there may exist distinct thin vortex fluxes which contribute
an identical phase to any Wilson loop they link. An example of this
was given already in section \ref{examse}; in the $SU(2)$ thin vortex
configuration, eq. (\ref{thinv}),
\begin{equation}
a_{\varphi } = \pm \frac{1}{r} T_3 \ , \ \ \ a_r =0
\end{equation}
both choices of sign lead to the same effect on a linked Wilson loop.
Both choices correspond to a Cartan chromomagnetic flux described by
$E_3 =2\pi $, cf. eq. (\ref{su2e}), but they differ in the direction, or 
orientation, of the flux, which is inverted when the sign is reversed.
\medskip

On the other hand, on a $Z(2)$ lattice, one cannot distinguish between
these two possibilities, since one formulates the theory directly in
terms of group elements. On a $Z(2)$ lattice, there is only one type
of center vortex flux, defined by plaquettes taking the value $(-1)$.
Information about the orientation of the flux is lost during lattice
center projection. For the purpose of evaluating Wilson loops, this
is immaterial; however, there may be other observables which are
sensitive to the orientation of vortex flux. An important example
of this, namely the Pontryagin index, will be discussed in detail in
section \ref{topsect}.
\medskip

Thus, to obtain a comprehensive description of infrared phenomena in
Yang-Mills theory, it is necessary to project onto a slightly more
general class of infrared effective degrees of freedom than the
essentially unoriented center vortices defined by lattice center
projection. Namely, vortex surfaces should additionally be associated
with an orientation, and should in general consist of patches of differing 
orientation. As will be shown in section \ref{topsect} in detail, the edges 
of these patches can be associated with Abelian magnetic monopole 
trajectories. The continuum center vortices introduced in the past sections 
do already include information about the orientation of magnetic flux. 
E.g., in the discussion after eq. (\ref{redefg}), when maximization over all 
hypersurfaces $\Sigma $ is called for, different orientations of $\Sigma $ 
can be used. Likewise, in the discussion after eq. (\ref{einfgc}), 
different orientations of $\partial \Sigma $, i.e. directions of flux 
in $a$, are distinguishable. However, vortex surface patches of alternating 
orientation, bounded by Abelian monopole currents, have hitherto not been 
allowed. On the contrary, the magnetic flux carried by thin vortex 
configurations $a$ has been assumed to be continuous throughout.
\medskip

The purpose of the the present discussion is merely to round out the
treatment of maximal center gauge fixing and center projection by pointing 
out how it can easily be generalized to yield vortex surfaces made up of 
oriented patches. More detailed properties of such surfaces, including how 
they determine the Pontryagin index, are discussed in section \ref{topsect}.
\medskip

One way of introducing patches of different orientation on the vortex 
surfaces is to include information on Abelian monopole degrees of
freedom into the gauge fixing and projection procedure. As mentioned
above, the trajectories of such monopoles on vortex surfaces define
the edges of the oriented patches. Such a procedure has already been 
defined and implemented in the lattice formulation, namely the 
``Indirect Maximal Center Gauge'' introduced in \cite{deb97} and further 
investigated in \cite{deb97aug}. This procedure 
is defined by initially transforming a given lattice Yang-Mills 
configuration to the maximal Abelian gauge and performing Abelian 
projection, which allows to extract Abelian monopole trajectories. 
In a subsequent step, one transforms the residual Abelian lattice 
configuration to the maximal center gauge and performs center projection, 
allowing the extraction of vortex surfaces. This two-step procedure, 
including two truncations of the theory by projection, is easily 
translated into the continuum formulation; the maximal Abelian gauge 
corresponds to the gauge fixing condition
\begin{equation}
[ \partial_{\mu } + A_{\mu }^{(n)} , A_{\mu }^{(ch)} ] = 0
\end{equation}
where $A_{\mu }^{(n)} $ denotes the color neutral (Cartan) components of 
the gauge field, and $A_{\mu }^{(ch)} $ denotes the color charged 
components. Fixing to this gauge induces Abelian monopoles in the
resulting field configurations. After projecting the gauge-fixed 
configurations onto their color neutral part (Abelian projection), the
maximal center gauge fixing procedure, as described in the past sections,
can be implemented analogously, and the associated vortices extracted.
\medskip

Note that the construction does not guarantee a priori that the monopoles 
are located on the vortex surfaces. However, monopoles have been found 
empirically to lie on the center vortices extracted via the indirect 
maximal center gauge in lattice experiments \cite{deb97aug}.
\medskip

On the other hand, it is not strictly necessary to resort to such a
two-step procedure, one of the steps being geared to extract Abelian
monopoles, and the other to extract center vortices. The coset gauge
transformations $\bar{G} $ in the maximal center gauge fixing functional
(\ref{mcgc}) in particular include gauge transformations which generate
Dirac strings\footnote{Note that the description in terms of Dirac strings,
used in the following for convenience, is equivalent to a description
in terms of Abelian monopoles. The two objects are strictly coupled.
Monopole trajectories define the edges of Dirac string world-sheets;
Dirac string world-sheets span areas circumscribed by monopole
trajectories, where the specific area associated with a monopole
trajectory can be modified by gauge transformations.}. Dirac strings,
which describe open two-dimensional world-sheets in four-dimensional
space-time, carry quantized magnetic flux such that they do not influence
any Wilson loop, in accordance with the fact that they are unobservable
pure gauge artefacts. E.g., in the case of $SU(2)$ color, a Dirac string
carries twice the magnetic flux of a thin vortex. Before discussing the
interplay between vortices and Dirac strings, consider the following example
for illustration. The $SU(2)$ pure gauge configuration
\begin{equation}
a^{dirac} = \bar{G} \partial \bar{G}^{\dagger } \ , \ 
\bar{G} =\left( \begin{array}{cc} 
\cos (\theta /2) & \sin (\theta /2) e^{-i\varphi } \\
-\sin (\theta /2) e^{i\varphi } & \cos (\theta /2)
\end{array} \right)
\end{equation}
where $(r,\theta ,\varphi )$ denotes the usual three-dimensional spherical
coordinates, and the configuration may be considered constant in the
further space-time coordinate, has the explicit components
\begin{eqnarray}
& a^{dirac}_{r} = 0 \ , \ \ \ \ \ \ \ \ \
a^{dirac}_{\theta } = \frac{1}{2r}
\left( \begin{array}{cc}
0 & -e^{-i\varphi } \\ e^{i\varphi } & 0
\end{array} \right) & \\
& a^{dirac}_{\varphi } = \frac{i}{r\sin \theta } \sin (\theta /2)
\left( \begin{array}{cc}
\sin (\theta /2) & \cos (\theta /2) e^{-i\varphi } \\
\cos (\theta /2) e^{i\varphi } & -\sin (\theta /2)
\end{array} \right) &
\end{eqnarray}
The full nonabelian field strength of this configuration can easily be
verified to vanish \cite{quadis}, except on the negative 
$z$-axis\footnote{The action of this pure gauge configuration of course 
must vanish in spite of the singularity on the negative $z$-axis, cf. the
pertinent comments in Appendix \ref{idappv}.}, i.e. for
$\theta =\pi $. There, the Dirac string is located. For 
$\theta \rightarrow \pi $, $a^{dirac}_{\varphi } $ can be approximated
by
\begin{equation}
a^{dirac}_{\varphi } \approx -\frac{1}{r\sin \theta } \cdot 2T_3
= 2a^{thin \ vortex}_{\varphi }
\end{equation}
and the flux through an infinitesimal Wilson loop encircling the Dirac
string therefore is
\begin{equation}
\Phi = \int_{0}^{2\pi } d\varphi \, r\sin \theta \, a^{dirac}_{\varphi }
= -4\pi T_3
\end{equation}
leading, upon exponentiation, to the trivial value $W=1$ for the Wilson 
loop. Furthermore, by considering the Abelian projection $\bar{a}^{dirac} $ 
of $a^{dirac} $,
\begin{equation}
\bar{a}^{dirac}_{\varphi } = \frac{-2\sin^{2} (\theta /2)}{r\sin \theta }
T_3 \ , \ \ \ \bar{a}^{dirac}_{\theta } = \bar{a}^{dirac}_{r} =0 \ ,
\end{equation}
one obtains in the associated Abelian magnetic field
\begin{equation}
\bar{B} = \partial \times \bar{a}^{dirac} = -\frac{1}{r^2 } e_r T_3
\label{dimono}
\end{equation}
a Dirac magnetic monopole located at the boundary of the Dirac string. Note 
that in the full field strength, this field is completely canceled by the 
nonabelian part \cite{quadis}, as mentioned above. Thus, magnetic 
monopoles which become visible in Abelian-projected configurations can 
be used as an alternative way of detecting, or describing, the Dirac 
strings induced by coset gauge fixing.
\medskip

The above properties imply that a Dirac string world-sheet which coincides
with a (oppositely oriented) patch of a thin $SU(2)$ vortex world-sheet 
simply reverses the magnetic flux associated with that patch, leaving all 
Wilson loops invariant. Nevertheless, in this way, though representing a 
pure gauge on its own, a Dirac string may change what information contained 
in a configuration is kept during continuum center (or vortex) projection,
i.e. it can be used to define a different truncation of the theory.
The physical content of the full configurations of course is not 
influenced by the Dirac string.
\medskip

It should be noted that for higher $SU(N)$ color groups, the 
superposition of a Dirac string on a vortex changes not only the
sign (i.e. the orientation), but in general also the type of vortex flux,
i.e. its direction in color space. E.g. for $SU(3)$, one in general 
additionally exchanges the center flux labels 
$(k=1) \leftrightarrow (k=2)$ in eq. (\ref{su3e}). This happens precisely 
in such a way as to leave all Wilson loops invariant, as it should be.
\medskip

Thus, due to the possibility of vortex surfaces being partially covered
by Dirac string world-sheets generated by coset transformations $\bar{G} $
in (\ref{mcgc}), vortex surfaces made up of patches of different
orientation are automatically generated during continuum center gauge
fixing. Vortex projection must merely be generalized to include this
additional information about the orientation of vortex magnetic flux
resulting from the Dirac world-sheets generated by coset gauge fixing.
\medskip

Finally, it should be noted that thick analogues of vortex
surfaces consisting of patches of different orientation have been
constructed in \cite{corn1},\cite{corn2}. Using such generalized objects
as approximating configurations $a$ in the gauge fixing condition
(\ref{mcgc}) provides another, alternative, way of extracting information
about vortex orientation from gauge field configurations.
\medskip

\section{Sketch of an Effective Vortex Theory}
\label{effthsec}
The center dominance observed for the string tension in lattice 
Monte Carlo calculations using the maximal center gauge 
\cite{deb97aug},\cite{giedt},\cite{deb97},\cite{tempv},\cite{tlang}
supports the notion that center vortices are the infrared 
degrees of freedom relevant for confinement in Yang-Mills theory. 
On a formal level, center or vortex dominance implies that the confinement 
properties can be adequately described by an effective vortex theory which 
results by appropriately integrating out all other Yang-Mills degrees of 
freedom. More specifically, in the abovementioned lattice calculations,
center dominance is obtained using the full lattice Yang-Mills dynamics,
i.e. the field configurations are sampled according to the full
Yang-Mills action. Center projection is only performed in the observable, 
i.e. the Wilson loop is evaluated using center-projected configurations
after maximal center gauge fixing. Thus, in principle, one could evaluate 
the expectation value of the Wilson loop using a $Z(N)$ effective action 
obtained after integrating out all coset $SU(N)/Z(N)$ degrees of freedom. 
In lattice calculations, this separation of the functional integrations 
is impractical, and the $Z(N)$ effective action is unknown. 
\medskip

Adopting a continuum description does not greatly alleviate this
problem, and therefore the following considerations concerning the
effective theory of vortices must remain mostly on a formal level.
The intention is to sketch what types of effects may arise in the
effective vortex theory particularly in the long-wavelength domain.
\medskip

To this end, it is necessary to implement a maximal center gauge
in the Yang-Mills functional integral. This gauge fixing can be
decomposed into two steps, as shown in section \ref{congsec}, namely
coset gauge fixing in an arbitrary fixed thin vortex configuration
$a$ and optimization of the thin vortex configuration. Coset
gauge fixing can be achieved in the standard way via the Fadeev-Popov
procedure, i.e. the gauge fixed partition function takes the form
\begin{equation}
Z = \int [DA] \, \Delta_{F} [A,a] \exp \left(
-S_{YM} [A] - pF^2 [A,a] \right)
\label{cofiz}
\end{equation}
where $F[A,a]$ denotes the gauge fixing condition (\ref{gaucon}),
$\Delta_{F} [A,a]$ is the standard Fadeev-Popov determinant corresponding
to this gauge fixing condition, and $p$ is a gauge fixing parameter
which should be sent to infinity in order to implement the coset
gauge fixing condition (\ref{gaucon}) exactly. Note that at this point
$a$ is an arbitrary, but fixed, external (thin vortex) field configuration.
In order to optimize the choice of thin vortex field $a$, it is useful
to further introduce into the functional integral the following 
representation of unity\footnote{Note that one could also forego
separating coset gauge fixing and vortex optimization, and instead
use the full continuum maximal center gauge fixing functional (\ref{mcgc})
to introduce a representation of unity into the Yang-Mills functional
integral analogously to (\ref{uniteinf}).},
\begin{eqnarray}
1 &=& \Delta [A] \int [D\Sigma ] \exp \left( 
-S_{GF} [A,a(\partial \Sigma ) ] \right) \label{uniteinf} \\
S_{GF} &=& -q \int d^D x \, g \left( \mbox{tr} (A-a)^2 \right)
\end{eqnarray}
cf. eq. (\ref{einfgc}). Here, the integration runs over all 
$(D-1)$-dimensional hypersurfaces $\Sigma $ in $D$ space-time dimensions,
and the specific realization $a(\partial \Sigma )$ of a thin vortex on
$\partial \Sigma $ can e.g. be given by (\ref{arbthv}). Furthermore,
the gauge fixing parameter $q$ should also be sent to infinity in order
for $e^{-S_{GF} } $ to be peaked at the optimal thin vortex configuration.
\medskip

Note that (\ref{uniteinf}), in order to be properly defined, at the very 
least requires some further gauge fixing due to the fact that the
integrand depends only on the boundaries $\partial \Sigma $ of the
hypersurfaces $\Sigma $ being integrated over. Note that it is by no
means clear that the integral over hypersurfaces $\Sigma $ can simply
be replaced by an integral over closed surfaces $S$; there may exist closed
surfaces $S$ which cannot be represented as $S=\partial \Sigma $ in terms of
some hypersurface $\Sigma $. A simple example is one of the coordinate
planes, e.g. the 1-2-plane, in a four-dimensional space-time hypercube
endowed with periodic boundary conditions, i.e. a torus. Such a plane
constitutes a closed surface due to the periodic boundary conditions,
but cannot be represented as the boundary of any three-volume $\Sigma $.
In general, the number of topologically inequivalent closed
two-dimensional surfaces which are not boundaries of three-volumes is
given by the dimension of the 2nd homology group $H_2 (M)$ of the
space-time manifold $M$ under consideration; this corresponds to the
2nd Betti number $b_2 = \mbox{dim} H_2 (M)$. On the four-dimensional
torus $T^4 $, one has $b_2 (T^4 ) =6$, corresponding precisely to the
six independent planes which are closed by the periodic boundary conditions.
\medskip

Formally, one can e.g. modify (\ref{uniteinf}) by writing
\begin{equation}
1 = \bar{\Delta } [A] \int [DS] \int [D\Sigma ] 
\delta [\Sigma - \Sigma_{min} [S] ]
\exp \left( -S_{GF} [A,a(S)] \right)
\label{uneinmod}
\end{equation}
\medskip
where $\Sigma_{min} [S]$ denotes the hypersurface of minimal volume
which has $S$ as its boundary, $\partial \Sigma_{min} =S$. If no
hypersurface $\Sigma $ exists such that $\partial \Sigma =S$, then
that $S$ does not contribute in (\ref{uneinmod}). Note that this
modification, leaving from the original integration over hypersurfaces
$\Sigma $ only an integration over closed $(D-2)$-dimensional surfaces $S$ 
in $D$ space-time dimensions, corresponds precisely to a fixing of the
center part of the gauge freedom on the lattice. Note also that this 
further gauge fixing in general leads to further measure factors, cf.
the modified measure $\bar{\Delta } [A]$ in (\ref{uneinmod}) as opposed
to $\Delta [A]$ in (\ref{uniteinf}).
\medskip

Inserting (\ref{uneinmod}) into (\ref{cofiz}), and assuming functional
integrations can be interchanged, one arrives at
\begin{eqnarray}
Z &=& \int [DS] e^{-S_{eff} [S] } \label{efvosi} \\
e^{-S_{eff} [S] } &=&
\int [D\Sigma ] \delta [\Sigma - \Sigma_{min} [S] ] \label{effsigt} \\
& & \ \ \ \ \ \ \ \cdot
\int [DA] \, \bar{\Delta } [A] \Delta_{F} [A,a] 
e^{-S_{YM} [A] - pF^2 [A,a] -S_{GF} [A,a] }
\nonumber
\end{eqnarray}
where $a\equiv a(S)$, as specified e.g. by (\ref{arbthv}).
The effective action given by (\ref{effsigt}) describes a theory of
$(D-2)$-dimensional closed vortex surfaces $S$ in $D$ space-time dimensions.
In the most general case, e.g. for the purpose of evaluating the
vortex-projected Pontryagin index, cf. next section, one will supplement
the surfaces $S$ with information about Dirac strings contained in the
coset gauge-fixed configurations $A$; this leads to patches of
different orientation  on $S$ as indicated in section \ref{altvog}.
Formally, the surfaces $S$ being integrated over in (\ref{efvosi})
will consist of oriented patches, and the integration over $A$ in
(\ref{effsigt}) for given $S$ will in turn be restricted to all $A$
containing the corresponding Dirac string configuration.
\medskip

Of course, the previous manipulations are at this stage purely formal.
Even to obtain the measure $\bar{\Delta } [A]$ in (\ref{effsigt}), one has
to solve a string theory in an arbitrary gauge background $A$ in four 
space-time dimensions, namely one has to carry out the integral over $S$ 
in (\ref{uneinmod}). However, in lower dimensions, an evaluation 
of (\ref{uneinmod}) may be feasible. E.g., in three space-time dimensions, 
the $S$ denote closed lines, and the integration over $S$ may be converted 
to a field theory by interpreting these lines as world-lines of particles.
\medskip

In order to nevertheless gain some insight into the structure of the
vortex effective action at least in the infrared limit, assume the
validity of the statement that the center projection vortices in
Yang-Mills theory give a rough localization of thick vortices present 
in the full Yang-Mills configurations. As discussed in section
\ref{examse}, this statement is corroborated by empirical findings
in lattice experiments \cite{deb97} and can be understood on the basis of 
the continuum formulation of the maximal center gauge developed in
section \ref{tcontlim}. However, it is a statement which can at
most be considered valid on length scales coarser than the typical
thickness $R$ of the aforementioned vortices in the full configurations
preferred by the Yang-Mills dynamics. At this point, $R$ is an
undetermined scale; it should ultimately be extracted by minimizing
the effective thick vortex action, as indicated further below.
\medskip

The above statement implies that the distribution of thin projection 
vortex configurations in the infrared regime is simply determined
by the distribution of thick vortices in Yang-Mills theory. In other
words, the infrared effective action of projection vortices can be
obtained by calculating the standard Yang-Mills effective action of
thick vortex configurations, using a thick vortex background gauge
as already briefly discussed in section \ref{altvog}. The ultraviolet
limit of validity of this effective action is given by the cutoff
$\Lambda = 1/R$.
\medskip

Furthermore, in such a long-wavelength approximation scheme, it is
consistent to evaluate the effective action in a gradient expansion,
i.e. in powers of $R\partial_{i} $, where 
$\partial_{i} \equiv \partial / \partial \sigma_{i} $ is the
derivative with respect to the parameters of the vortex surface.
Below, this effective action will be given in the classical limit,
i.e. to zeroth order in a loop expansion. This will already exhibit
the relevant terms, with higher orders in the loop expansion only
renormalizing the coefficients. The calculation is carried out explicitly
in Appendix \ref{idappv} using, for convenience, ultraviolet-regulated 
ideal vortex configurations. More precisely, the $\delta $-function in the 
defining expression (\ref{idvorc}) for the ideal vortex 
${\cal A} (k,\Sigma ,x)$ is regularized by means of the cutoff 
$\Lambda = 1/R$, cf. eq. (\ref{regidv}). Note that
such an object represents a bona fide thick vortex; the special
character of ideal vortices as singular continuum representations
of excitations on a $Z(N)$ lattice is blurred when the ideal vortex
singularities are regularized. After regularization, there is no
qualitative distinction between an initially thin or an initially ideal
vortex; these can all be viewed as continuum coset configurations,
namely, thick vortices.
\medskip

Inserting such a thickened ideal vortex ${\cal A} $ into the classical
Yang-Mills action,
\begin{equation}
S [ {\cal A} ] = - \frac{1}{2 g^2_0} \int \mbox{tr} 
\left( F [ {\cal A} ] F [ {\cal A} ] \right)
\end{equation}
one obtains the gradient expansion (see Appendix \ref{idappv})
\begin{eqnarray}
S_{YM} [ {\cal A} ] = \alpha \int_{\partial \Sigma } d^2 \sigma \sqrt{g} 
&+& \beta \int_{\partial \Sigma } d^2 \sigma \sqrt{g}  K^A_{ai} K^A_{bj} 
g^{ab} g^{ij} \nonumber \\
&+& \gamma \int d^2 \sigma \sqrt{g} g^{ab} \partial_{a} \partial_{b} 
\ln \sqrt{g} \ .
\label{effclaa}
\end{eqnarray}
Here,
\begin{equation}
g = \det g_{ab} \ , \ g_{ab} = \partial_{a} x^{\mu } \partial_{b} x^{\mu } 
\end{equation}
denotes the metric on the vortex sheet and $K^A_{ab} $ are the extrinsic 
curvature coefficients which are defined in Appendix \ref{idappv},
eq. (\ref{kabdec}). Furthermore, the coefficients in eq. (\ref{effclaa}) 
are given for $SU(2)$ vortices by 
(using $\mbox{tr} E^2 = -\frac{1}{2} (2\pi )^2 $, cf. eq. (\ref{su2e}) )
\begin{equation}
\alpha = \frac{\pi }{g^2_0 R^2 } \ ,
\ \ \ \ \ \ \beta = -\frac{\pi }{2 g^2_0 } \ ,
\ \ \ \ \ \ \gamma = \frac{\pi }{4 g^2_0 } \ .
\label{effacoe}
\end{equation}
One thus obtains in leading and next-to-leading order the Nambu-Goto
action and curvature terms for the vortex surfaces\footnote{The second
term on the r.h.s. of eq. (\ref{effclaa}) has been discussed in 
\cite{polyea},\cite{kleinea}.}. The coefficients (\ref{effacoe}) will be 
renormalized through quantum effects embodied by higher orders in the loop 
expansion. In particular, renormalization will generate a physical scale in 
the coefficients in the standard fashion, which will allow to relate the 
thickness $R$ to physically measurable quantities. This will make it 
possible to self-consistently determine the preferred physical vortex 
thickness $R$ through a minimization of the effective action for given 
vortex magnetic flux; similar considerations were already carried out in
the framework of the Copenhagen vacuum \cite{spag}. Such more detailed
approximations, e.g. an evaluation of the one-loop corrections,
are deferred to future work.
\medskip

It should however be emphasized that, though the explicit form of the 
coefficients of the various terms in the effective vortex action certainly 
depends on the approximation involved, the general structure of the 
gradient expansion of the vortex effective action is uniquely determined 
by Poincar\'{e}, gauge, and reparametrization invariance. In a forthcoming 
paper, the authors will use an action reminiscent of (\ref{effclaa}) 
with adjustable coefficients as a starting point for a phenomenological
vortex model of infrared Yang-Mills physics.
\medskip

\section{Topology of Center Vortices}
\label{topsect}
In the previous section, a rough sketch of a reduction of Yang-Mills 
theory to an effective vortex theory was given, by implementing a
continuum version of the maximal center gauge, and by subsequently 
integrating out the Yang-Mills fields. Center-projected (or, more 
appropriately, vortex-projected) observables can in principle now be 
evaluated directly in terms of the vortex theory, which contains 
exclusively $(D-2)$-dimensional vortex surface degrees of freedom in 
$D$-dimensional space-time.
\medskip

There is an essential difference between the center vortices on the lattice 
and in the continuum theory. After center projection on the lattice, the
direction of the magnetic flux of the vortex is lost, while in the continuum 
theory, the center vortices are given by oriented (patches of) surfaces, 
where the flux direction is defined by the orientation. This was already
discussed to some extent in section \ref{altvog}. The orientation of the 
vortices, however, is crucial for their topological properties, as will 
be shown below by evaluating the Pontryagin index
\begin{eqnarray}
\nu [A] = \frac{1}{32\pi^{2} } \int d^4 x \, 
\tilde{F}^{a}_{\mu \nu } F^{a}_{\mu \nu } = 
- \frac{1}{16\pi^2} \int d^4 x \, \mbox{tr} \, 
\tilde{F}_{\mu \nu } F_{\mu \nu } 
\label{nudefs}
\end{eqnarray}
using the ideal vortex representation (\ref{idvorc}).
\medskip

It should be emphasized that the treatment in this section, the goal of
which is to define the vortex-projected Pontryagin index, in some
ways reverses the logic of the past sections. Whereas the continuum
maximal center gauge and also the vortex action were defined via the
corresponding lattice expressions in the limit of vanishing lattice
spacing, the Pontryagin index calls for a different treatment. It
is initially defined for smooth continuum gauge field configurations,
and useful lattice discretizations of the Pontryagin index are
notoriously cumbersome. In fact, the maximal center gauge exacerbates
the problem, since all lattice definitions at some point must assume
the links of a lattice configuration to be reasonably close to
unity. On the other hand, the maximal center gauge precisely transforms
configurations into a form where their links strongly deviate from unity;
namely, it introduces ideal vortices even into initially smooth
configurations. Therefore, to properly define the vortex-projected 
Pontryagin index, it is necessary to start from the continuum
expression, understanding the Pontryagin index for thin (or, 
gauge-equivalently, ideal) vortices to correspond to the thin limit of
the Pontryagin index of thickened vortices, for which it is well-defined.
This limit is smooth, as is demonstrated in an example in
Appendix \ref{appc}; namely, the thick vortex profile function
manifestly cancels. Note that this concept of vortex projection
is not directly connected with lattice center projection anymore,
since it is based on the thin limit of thickened (i.e. coset)
configurations. In the following, the Pontryagin index will be 
evaluated for (thickened) ideal vortex configurations. The aforementioned 
example in Appendix \ref{appc} on the other hand treats a particular 
configuration of explicitly thick vortices, containing also the limiting 
case of thin vortices.
\medskip

The field strength of an ideal center vortex ${\cal A}$ is given by 
\begin{eqnarray}
{\cal F}_{\mu\nu} (k,x,S=\partial \Sigma ) = F_{\mu \nu } [ {\cal A} ] =
\partial_{\mu} {\cal A}_{\nu} (k,\Sigma ,x) 
- \partial_{\nu} {\cal A}_{\mu} (k,\Sigma ,x)
\label{calfca}
\end{eqnarray}
Straightforward evaluation (see Appendix \ref{idappv}) yields for
the field strength\footnote{In the context of superfluid Helium, this 
quantity is referred to as the vorticity tensor.} on a vortex
surface $S$
\begin{equation}
{\cal F}_{\mu\nu}(k,x,S) = E(k) \int_{S} d^2 \tilde{\sigma}_{\mu\nu} 
\delta^4 (x - \bar{x} (\sigma) ) \ .
\label{cloinn}
\end{equation}
One can decompose a general vortex configuration into components of
different center flux labeled by the integer $k$, where
$k=1,\ldots ,N-1 $ for $SU(N)$ color,
\begin{equation}
S=\bigcup_{k} S_k \ , \ \ \ {\cal F}_{\mu \nu } (x,S) =
\sum_{k} {\cal F}_{\mu \nu } (x,S_k ) \ ;
\end{equation}
Inserting this, and eq. (\ref{cloinn}), into (\ref{nudefs}),
one finds for the Pontryagin index 
\begin{equation}
\nu [{\cal A}] = -\frac{1}{8\pi^{2} } \sum_{k,k^{\prime } } 
I(S_k ,S_{k^{\prime } } ) \mbox{tr} ( E(k) E(k^{\prime } ) )
\label{yidres}
\end{equation}
where
\begin{equation}
I(S_1 , S_2 ) = \frac{1}{2} \int_{S_1 } d^2 \tilde{\sigma }_{\mu \nu } 
\int_{S_2 } d^2 \sigma^{\prime }_{\mu \nu } 
\delta^{4} (\bar{x} (\sigma ) - \bar{x} (\sigma^{\prime } ) )
\label{isndef}
\end{equation}
is the (oriented) intersection number between two two-dimensional 
surfaces $S_1, S_2$ in $D = 4$ space-time dimensions. Note that the 
intersection number $I(S_1 ,S_2 )$ is also defined for open surfaces 
$S_1 ,S_2 $, and that, in $D=4$ dimensions, $I(S_1 ,S_2 ) = I(S_2 ,S_1 )$.
Also, if $S_1 $ and $S_2 $ intersect in a single point, then
$I(S_1 ,S_2 ) =\pm 1$, where the sign depends on the relative orientation
of $S_1 $ and $S_2 $. Note furthermore that, in the case $k=k^{\prime } $
in (\ref{yidres}), $I(S_k ,S_k )$ counts each isolated self-intersection
point of $S_k $ twice. This is easily seen by further decomposing $S_k $
into two components $S_k^1 $ and $S_k^2 $ such that neither of the two
components self-intersect at the point in question. Then,
\begin{equation}
I(S_k ,S_k ) = I(S_k^1 ,S_k^1 ) + I(S_k^2 ,S_k^2 ) +2I(S_k^1 ,S_k^2 )
\end{equation}
where the final term represents the contribution from the 
self-intersection point of $S_k $ under consideration.
\medskip

It should be remarked that this discussion of the self-intersection 
number $I(S,S)$ includes only the contributions from transversal 
intersection points, i.e. points where 
$\bar{x}_{\mu } (\sigma ) = \bar{x}_{\mu } (\sigma^{\prime } ) $ with
$\sigma \neq \sigma^{\prime } $, but leaves out the contribution from
the coincidence points
$\bar{x}_{\mu } (\sigma ) = \bar{x}_{\mu } (\sigma^{\prime } ) $
with $\sigma =\sigma^{\prime } $. A more thorough discussion of the
self-intersection number $I(S,S)$, which is given in Appendix \ref{intseca}, 
does, however, not alter the conclusions on the Pontryagin index reached 
here.
\medskip

Thus, the Pontryagin index of a vortex surface configuration is built up
from its self-intersection points, each of which enters with a
contribution $\pm 2\cdot \mbox{tr} ( E(k) E(k^{\prime } ) ) /8\pi^{2} $ 
into the Pontryagin index according to (\ref{yidres}). The integers 
$k,k^{\prime } $ label the center flux carried by the two patches of the 
vortex surface configuration intersecting at the point in question; the 
sign depends on the relative orientation of the patches. In view of 
eq. (\ref{su2e}), this implies specifically for $SU(2)$ color a 
contribution $\pm 1/2$ from each self-intersection point; on the other 
hand, for $SU(3)$ color, each self-intersection point contributes 
$\pm 1/3$ or $\pm 2/3$ to the Pontryagin index, cf. eqs. (\ref{scprd})
and (\ref{scpro}). This is consistent with the dependence on the
number of colors noted by J.~M.~Cornwall in \cite{corn1}.
\medskip

In view of this result, the reader might worry at this point about the
integer-valuedness of the Pontryagin index on $R^4 $ (silently 
compactified to $S^4 $). This property cannot go lost during vortex
projection, for the very simple reason that vortices can be explicitly
represented as gauge field configurations and therefore inherit all
properties known to be true for generic gauge fields. Nevertheless,
it is instructive to remark how the integer-valuedness comes about in
more detail in a simple case, namely $SU(2)$ vortices on $R^4 $.
It is due to the fact well-known in topology\footnote{The authors thank
B.~Leeb for a discussion on this point.} that the number of intersection
points of two closed two-dimensional surfaces in $R^4 $ is even. This
also implies that the number of self-intersection points of a closed 
surface is even, because the self-intersection number is defined by 
simply intersecting the surface with another surface infinitesimally 
displaced from it (i.e. one considers a framing of the surface, 
cf. Appendix \ref{intseca}). Thus, while each self-intersection point of 
an $SU(2)$ vortex surface configuration gives a contribution $\pm 1/2$ to 
the Pontryagin index, the number of such contributions is even, and the
Pontryagin index integer-valued. 
\medskip

In the case of higher $SU(N)$ gauge groups, the argument becomes more 
complicated, since the surfaces may branch and also due to the fact that 
superimposed Dirac strings in general also modify the type of vortex flux, 
i.e. its direction in color space. Further complications arise on different 
space-time manifolds. For instance, on a four-cube with periodic boundary 
conditions, i.e. a torus, one might think of generating a single 
intersection point between two surfaces by choosing e.g. the 1-2 plane and 
the 3-4 plane, which are closed due to the periodic boundary conditions. 
However, this example is invalid, since vortex surfaces must be 
representable as surfaces of three-volumes, which is not the case for the 
aforementioned planes.
\medskip

\begin{figure}[ht]

\vspace{-7.5cm}

\centerline{
\epsfysize=15cm
\epsffile{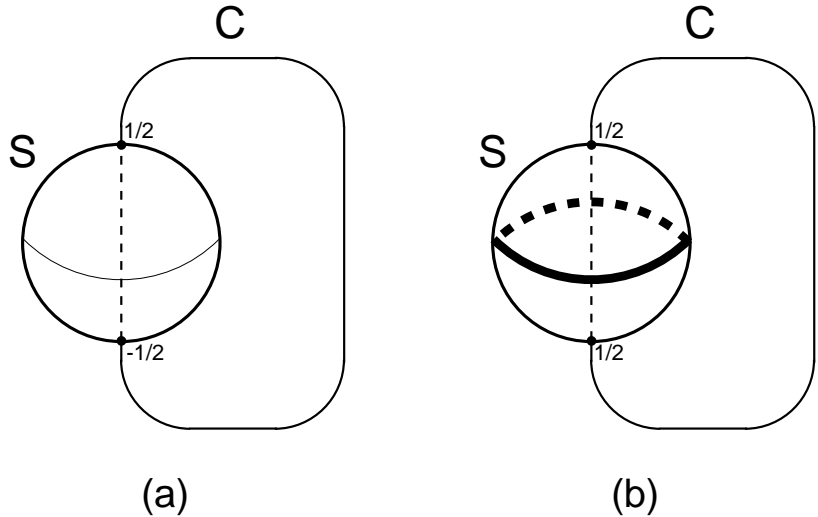}
}
\caption{Three-dimensional slice of two intersecting SU(2) vortex surfaces
in four space-time dimensions (see text). The thin line around the equator 
of the sphere $S$ in (a) is merely to guide the eye in identifying $S$ as 
a sphere. At the intersection points of $S$ and $C$, contributions to the 
Pontryagin index of modulus $1/2$ arise. However, in the case of globally 
oriented surfaces (a), these contributions cancel. On the other hand, in 
the case that vortex surfaces are not globally oriented, but consist of 
patches of different orientation (b), the contributions may add, giving 
rise to a nonvanishing Pontryagin index. Boundaries of such patches are 
tagged by Abelian monopoles (thick line around the equator in (b)).}
\label{fig3}
\end{figure}

The final point to be discussed is the fact, also well-known in topology,
that the self-intersection number of closed, {\em globally oriented}
two-dimensional surfaces in $R^4 $ not only is a multiple of four
(remember that each intersection point is counted doubly, and there
are an even number of intersection points for closed surfaces), but
actually vanishes. This implies that the Pontryagin index vanishes
for globally oriented vortex surfaces; conversely, therefore,
non-orientedness of the surfaces is crucial for generating a
non-vanishing topological winding number. To illustrate this, consider the 
following three-dimensional slice of two intersecting $SU(2)$ vortex 
surfaces in $D=4$ dimensions, cf. Fig.~\ref{fig3}. Let the one vortex be 
located entirely within the three-dimensional slice of space-time under 
consideration; it is therefore visible as a closed surface, namely the 
sphere $S$ in Fig.~\ref{fig3}. On the other hand, let the other vortex 
extend into the space-time dimension not displayed in Fig.~\ref{fig3}; it 
is then visible as a closed loop $C$ after slicing. Let $C$ intersect $S$ 
at two points, chosen in Fig.~\ref{fig3} as the poles of $S$.
\medskip

If $S$ and $C$ are both globally oriented, as in Fig.~\ref{fig3} (a), 
their intersection number (in $D=3$)
\begin{equation}
I(S,C) = \int_{S} d^2 \tilde{\sigma}_i \int_{C} dx_i \delta^3 (x -
\bar{x} (\sigma))
\end{equation}
vanishes, since the two intersection points occur with opposite relative
orientation between the intersecting manifolds. Thus, a non-zero 
intersection number requires at least one of the two intersecting 
closed manifolds to be not globally oriented. Globally non-oriented 
hypersurfaces consist nevertheless of oriented patches. Assume for 
example the sphere $S$ in Fig.~\ref{fig3} (b) to consist of two 
hemispheres of opposite orientation. E.g., the southern hemisphere may 
be covered by a Dirac string world-sheet, simply reversing (in the case 
of $SU(2)$ color) the orientation associated with that hemisphere. The 
boundary of a Dirac string world-sheet is complemented by a monopole in 
the Abelian-projected configuration, cf. eq. (\ref{dimono}). Thus, in
Fig.~\ref{fig3} (b), one can associate an Abelian monopole current with 
the equator (thick line). Further below, it will be shown more generally
how boundaries of oriented patches on vortex surfaces imply monopole
currents.
\medskip

Now, in the presence of the monopole current at the equator in 
Fig.~\ref{fig3} (b), implying a reversal in orientation of, say, the 
southern hemisphere, the contributions to the Pontryagin index from the two 
intersection points no longer cancel, but add coherently, giving a 
Pontryagin index of modulus $|\nu | =1$. Thus, the depicted configuration 
represents an analogue of an instanton in the sense that it will generate 
a zero mode in the Dirac operator according to the index theorem.
The authors speculate that standard smoothing procedures used on
the lattice to detect instantons would transmute this configuration
into an instanton, since smoothing by construction alters configurations
such as to lower the Yang-Mills action. After smoothing, it is impossible
to discern what type of excitation precisely carried Pontryagin index
in the original configuration. Note that this picture of the generation
of topological winding number also correlates well with the observation
\cite{monumi1},\cite{monumi2} that instantons are encircled by monopole 
loops in the maximal Abelian gauge. It is precisely the monopole loop 
which tags the change in orientation in the vortex surface (the sphere $S$ 
in Fig.~\ref{fig3} (b)) between the two intersection points carrying the 
topological density, causing their contributions to the Pontryagin index 
to add up and generate the ``pre-instanton''.
\medskip

To see the emergence of the magnetic monopole loops at the boundaries
of vortex surface patches in more detail, consider an isolated patch;
this now represents an open two-dimensional surface sheet $S$ with boundary 
$\partial S = C$. To this surface $S$, one can relate a magnetic flux sheet 
by associating with it the field strength (see above)
\begin{equation}
{\cal F}_{\mu \nu } (x,S) = E \int_{S} d^2 \tilde{\sigma }_{\mu \nu } 
\delta^{4} (x - \bar{x} (\sigma ) ) \ .
\label{assofs}
\end{equation}
For a closed surface $S=\partial \Sigma $, this expression agrees
with the one for the ideal vortex ${\cal A} (\Sigma ,x)$ given in
(\ref{cloinn}). By Stokes' theorem, one finds that a conserved (Abelian) 
magnetic current 
\begin{equation}
j^{m}_{\mu } (x,C) = \partial_{\nu } \tilde{ {\cal F} }_{\mu \nu } = 
E \oint_{\partial S =C} dy_{\mu } \delta^{4} (x-y) \ , \ \ \ \ \ \
\partial_{\mu } j_{\mu }^{m} (x,C) = 0
\end{equation}
must flow at the boundary $\partial S = C$ of the magnetic sheet $S$.
Note that the direction of the magnetic current on $\partial S = C$ defines 
an orientation of the sheet $S$ and vice versa. This monopole current is
generated by a magnetic point charge $q^m $
\begin{equation}
E q^m = \int_{V} d\tilde{\sigma }_{\mu } j^{m}_{\mu } (x,C)
= E I(V,C)
\label{magq}
\end{equation}
with $V$ denoting the three-volume dual to the loop $C$.
In view of $\exp (-E) = Z$, eq. (\ref{magq}) yields for $q^m $ only a
fraction of the elementary magnetic charge dictated by the Dirac 
quantization condition. This is as it should be, since the vortex
sheet $S$ only carries a fraction of the flux which a Dirac string
carries. Only when glueing the open vortex patches with their boundary
monopole loops back together, should one take care to obtain current 
loops of proper magnetic point charges satisfying the Dirac quantization 
condition. In this way, non-oriented closed magnetic vortex sheets are
constructed, consisting of oriented surface pieces joined by magnetic 
monopole current loops\footnote{If one glues together two patches such that 
the surface orientation does not change across the boundary, the monopole 
currents at the boundary precisely cancel.}. Such a non-oriented closed 
magnetic sheet defines a center vortex which still gives the same 
contribution to a Wilson loop as the corresponding oriented vortex 
(in the absence of a monopole loop), cf. eq. (\ref{wilsid}). Thus, for the 
confinement properties, measured by the Wilson loop, the orientation of a 
vortex sheet, and hence the magnetic monopole currents, are irrelevant. 
The Abelian magnetic monopole currents are however necessary in order to 
generate a non-vanishing Pontryagin index for vortex configurations. It is 
in fact easy to see that the Pontryagin index of the vortices can be 
determined from the Abelian magnetic monopoles alone. Indeed, inserting 
(\ref{calfca}) into (\ref{nudefs}), after a partial integration the 
Pontryagin index of an ideal vortex sheet becomes
\begin{eqnarray}
\nu &=& -\frac{1}{8\pi^{2} } \int_{\cal M} \partial_{\mu } \, \mbox{tr} \,
({\cal A}_{\nu } \tilde{\cal F}_{\mu \nu } ) +
\frac{1}{8\pi^{2} } \int_{\cal M} \, \mbox{tr} \,
{\cal A}_{\nu } \partial_{\mu } \tilde{\cal F}_{\mu \nu } \\
&=& -\frac{1}{8\pi^{2} } \int_{\partial {\cal M} } d\sigma_{\mu }
\mbox{tr} \, ({\cal A}_{\nu } \tilde{\cal F}_{\mu \nu } ) +
\frac{1}{8\pi^{2} } \int_{\cal M} \, \mbox{tr} \,
{\cal A}_{\nu } j_{\nu }^{m} (x,C)
\end{eqnarray}
where Gau\ss ' law was used in the first term and the definition of the
magnetic current in the second term. The second term obviously vanishes
in the absence of magnetic monopoles. Furthermore, the surface term
vanishes unless the integrand contains singularities which give rise to
internal surfaces wrapping the singularities. Such internal surfaces 
precisely arise in the presence of magnetic charges \cite{mmpoly} which 
may be magnetic monopoles or more extended magnetic charge distributions 
such as line or surface charges. Thus, a non-zero Pontryagin index requires
the existence of magnetic charges in Abelian-projected configurations. 
This is consistent with the findings in \cite{mmpoly}.
\medskip

\section{Outlook}
\label{outlsec}
In this work, the continuum analogues of the maximal center gauge and
center projection in lattice Yang-Mills theory were constructed. This shed
new light on the meaning of the procedure on the lattice and led to
a sketch of an effective vortex theory in the continuum. Also the manner
in which center vortex configurations generate the Pontryagin index was
clarified; the latter is built up from self-intersections of the vortex
network, where it is crucial that the vortex surfaces be globally 
non-oriented. Since these developments have already been discussed
at length in the corresponding sections above, it suffices here to
mention some open issues which require further investigation.
\medskip

With regard to the maximal center gauge, it is necessary to conduct
lattice experiments using alternative gauge fixing functions $\bar{g} $,
cf. eq. (\ref{mcgg}), and to test whether the physical conclusions 
reached using the direct maximal center gauge remain unaffected. If it is 
true that maximal center gauges provide a rough localization of thick 
vortices present in full lattice configurations, then long-distance 
properties should be robust with respect to a variation of $\bar{g} $; 
fluctuations of the center projection vortices within the thick vortex 
profile are of course gauge-dependent, as evidenced in section \ref{examse}.
An issue which is of particular interest in this context is the 
occurrence of Gribov copies which do not correspond to the absolute
maximum of the center gauge fixing functional. In \cite{tomuns},
the possibility is raised that the true gauge fixing image selected in
particular by the direct maximal center gauge contains less vortices than
found in previous studies, and no vortex confinement. As discussed in
sections \ref{revthsec} and \ref{dmcgsec}, the direct maximal center gauge 
in the continuum limit in fact always selects a trivial gauge fixing image 
devoid of center vortices. The observations of \cite{tomuns} may be a
signature of the specific problems of this particular gauge. On the other
hand, it is straightforward to formulate alternative gauge fixing
functions $\bar{g} $ which prevent these problems. It would be interesting
to repeat the considerations of \cite{tomuns} using such a gauge.
\medskip

Concerning the issue of an effective vortex theory, it was already mentioned 
in section \ref{effthsec} that the authors plan to report on their numerical 
study of a phenomenological random surface model for vortices in a 
forthcoming paper. On the other hand, from an analytical point of view,
it would presumably be helpful to start with lower-dimensional models.
E.g., it would be worthwhile to study whether the maximal center gauge
allows a solution of Yang-Mills theory in $1+1$ space-time dimensions;
there, the effective vortex theory takes the form of a classical
statistical mechanics of vortex points.
\medskip

Finally, it is necessary to investigate whether center vortices do in
fact fully capture the physics of the Pontryagin index; i.e. one
should test whether the vortex-projected Pontryagin index displays
vortex dominance in analogy to the center dominance observed for the
string tension. While it has been shown in this work how vortices
generate a nontrivial topological winding number, and the converse
experiment in \cite{forcrand} has shown that an ensemble devoid of
center vortices is concentrated in the trivial topological sector,
the question of vortex dominance is more stringent. In this context, it 
is worth mentioning the analogous result from Abelian-projected theories: 
In the maximal Abelian gauge, one finds monopole dominance in the string 
tension \cite{cdomma2}, but not in the Pontryagin index; on the other hand,
in the Polyakov gauge, the monopole dominance in the string tension is much
less pronounced, but the Pontryagin index can be reconstructed exactly
from the monopole content of the gauge-fixed configurations alone 
\cite{mmpoly}.
\medskip

In connection with the issue of the Pontryagin index, of course also
a more detailed study of chiral symmetry breaking by a vortex background
is called for. Vortex configurations which give rise to a non-vanishing
Pontryagin index must, by the index theorem, give rise to zero modes
of the Dirac operator. It would e.g. be interesting to explicitly
construct such zero modes for some relevant simple vortex configurations.
In the context of instanton models, the zero modes induced by the
instantons give in fact the dominant contribution to the chiral condensate.
The question arises whether the chiral condensate displays vortex
dominance (again, the converse experiment \cite{forcrand} yields zero
condensate in the absence of vortices), and, if so, how it behaves
at finite temperatures, in particular, at the deconfinement transition.
\medskip

\section*{Acknowledgements}
Discussions with K.~Langfeld and B.~Leeb are gratefully acknowledged.
\medskip

\begin{appendix}

\section{Properties of center vortices}
\subsection{The ideal center vortex}
\label{idappv}
In the following, the field strength $F_{\mu \nu } [ {\cal A} ] $ of the
ideal vortex field $ {\cal A} (k,\Sigma ,x)$ is calculated. For this
purpose, it is convenient to evaluate first its dual,
\begin{equation}
\tilde{\cal F}_{\mu \nu } =\frac{1}{2} \epsilon_{\mu \nu \alpha \beta }
F_{\alpha \beta } [ {\cal A} ] =
\epsilon_{\mu \nu \alpha \beta } \partial_{\alpha }
{\cal A}_{\beta } (k,\Sigma ,x)
\end{equation}
Inserting here the explicit representation for 
$ {\cal A}_{\beta } (k,\Sigma ,x)$, cf. eq. (\ref{idvorc}), one obtains
\begin{eqnarray}
\tilde{\cal F}_{\mu \nu } (k,\partial \Sigma ,x) &=&
E(k) \epsilon_{\mu \nu \kappa \lambda } \int_{\Sigma }
d^{D-1} \tilde{\sigma }_{\lambda } \partial_{\kappa }^{x}
\delta^{D} (x-\bar{x} (\sigma )) \\
&=& E(k) \frac{1}{3!} \epsilon_{\mu \nu \kappa \lambda }
\epsilon_{\lambda \alpha \rho \gamma } 
\int d^{D-1} \sigma_{\alpha \rho \gamma } \partial_{\kappa }^{x}
\delta^{D} (x-\bar{x} (\sigma )) \\
&=& E(k) \int_{\Sigma } d^{D-1} \sigma_{\kappa \mu \nu }
\partial_{\kappa }^{\bar{x} } \delta^{D} (x-\bar{x} (\sigma ))
\end{eqnarray}
Application of Stokes' theorem yields
\begin{equation}
\tilde{\cal F}_{\mu \nu } (k,\partial \Sigma ,x) =
E(k) \int_{\partial \Sigma } d^{D-2} \sigma_{\mu \nu }
\delta^{D} (x-\bar{x} (\sigma )) \ .
\label{fschid}
\end{equation}
Taking the dual of this equation yields the desired representation 
(\ref{cloinn}).
\medskip

With the vortex field strength eq. (\ref{fschid}) the Yang-Mills action 
of an ideal vortex in $D=4$ space-time dimensions is given (up to an 
unimportant numerical factor) by 
\begin{equation}
\bar{S} = \int_{\partial \Sigma } d^2 \sigma_{\mu \nu } 
\int_{\partial \Sigma } d^2 \sigma^{\prime }_{\mu \nu } 
\delta^{4} \left( \bar{x} (\sigma ) - \bar{x} (\sigma^{\prime } ) \right) \ .
\label{G1}
\end{equation}
Obviously this action is divergent and calls for regularization. 
Regularization will be achieved by chopping off the high frequencies using
an ultraviolet cutoff $\Lambda$; this amounts to replacing the 
$\delta$-function in equation (\ref{G1}) by the regularized version
\begin{equation}
\delta (x) = \frac{\Lambda }{\sqrt{2 \pi}} e^{-\frac{\Lambda^2 }{2} x^2 } 
\ .
\label{regidv}
\end{equation}
Before proceeding, a comment is in order concerning the thin limit
$\Lambda \rightarrow \infty $. As long as $\Lambda $ is arbitrarily
large, but fixed with respect to the ultraviolet regulator of the
theory, say the inverse lattice spacing, it is legitimate to 
evaluate the action using the continuum Yang-Mills expression
(\ref{G1}). However, if the vortex is meant to be truly thin, i.e. if
its entire magnetic flux is concentrated on one lattice plaquette,
then one must use the lattice plaquette expression for the action;
otherwise, one loses track of the compact character of the gauge group.
E.g., if one exactly superimposes two truly thin $SU(2)$ ideal vortices on 
top of each other (this corresponds to a closed Dirac string), then naive
application of the continuum Yang-Mills expression for the action
would lead to a value diverging like four times the action of a
single vortex. On the other hand, all plaquettes encircling this
Dirac string take the value $(+1)$, and its action therefore
vanishes, as it should\footnote{In the case of the Dirac string, these
subtleties can be circumvented by dividing the space-time manifold into
coordinate patches as in the Wu-Yang construction.}.
\medskip

Keeping this caveat in mind, the regularized action can be evaluated 
in a gradient expansion. Note that similar calculations have been performed 
in \cite{orsato},\cite{satoor} for a massive propagator. 
Since the action receives contributions only from 
$\bar{x} (\sigma) \simeq \bar{x} (\sigma')$, the integrand 
(\ref{G1}) can be expanded in powers of $z = \sigma - \sigma'$. 
Defining $s = \frac{\sigma + \sigma'}{2}$ and
expanding (where $\partial_{a} = \partial /\partial \sigma_{a} $),
\begin{eqnarray}
\bar{x} (\sigma) & = & \bar{x} (s) + \frac{z^a}{2} \partial_a \bar{x} (s) +
\cdots \nonumber\\
\bar{x} (\sigma') & = & \bar{x} (s) - \frac{z^a}{2} \partial_a \bar{x}
(s)\nonumber\\
\Sigma_{\mu \nu} (\sigma) & = & \Sigma_{\mu \nu} (s) + \frac{z^a}{2} \partial_a
\Sigma_{\mu \nu} (s) + \frac{1}{2} \frac{z^a}{2} \frac{z^b}{2} \partial_a
\partial_b \Sigma_{\mu \nu} (s) + \cdots \nonumber\\
\Sigma_{\mu \nu} (\sigma') & = & \Sigma_{\mu \nu} (s) - \frac{z^a}{2} \partial_a
\Sigma_{\mu \nu} (s) + \frac{1}{2} \frac{z^a}{2} \frac{z^b}{2} \partial_a
\partial_b \Sigma_{\mu \nu} (s) + \cdots \ ,
\end{eqnarray}
one obtains
\begin{eqnarray}
\label{G4}
\bar{S} & = & \left( \frac{\Lambda}{\sqrt{4 \pi}} \right)^{4} 
\int_{\partial \Sigma } d^2 \sigma \int_{-\infty }^{\infty } d^2 z \, 
e^{-\frac{\Lambda^{2} }{2} z^a g_{ab} (\sigma) z^b}
\left[ (\Sigma_{\mu \nu} (\sigma))^2 \right. \nonumber\\
& & \left. - \frac{z^a}{2} \frac{z^b}{2} \partial_a
\Sigma_{\mu \nu} (\sigma) \partial_b \Sigma_{\mu \nu} (\sigma) + \frac{z^a}{2}
\frac{z^b}{2} \Sigma_{\mu \nu} (\sigma) 
\partial_a \partial_b \Sigma_{\mu \nu}  (\sigma) 
\right] \ ,
\end{eqnarray}
where only terms up to order $z^2$ in both the exponent and
preexponent have been kept. Furthermore, the range of integration over $z$ 
has been extended to $\pm \infty $, since the main contribution to the 
integral comes from the region $z \approx 0$. Carrying out the Gaussian 
integration over $z$, one obtains
\begin{equation}
\bar{S} = \left( \frac{\Lambda }{\sqrt{2 \pi}} \right)^{2} 
\int_{\partial \Sigma } d^2 \sigma
\left[2 \sqrt{g} + \frac{1}{\Lambda^2} \left( \partial_a \Sigma_{\mu \nu}
\partial_b \Sigma_{\mu \nu} -  \frac{1}{2} \partial_a \partial_b g \right)
\frac{\partial}{\partial g_{ab}} \frac{1}{\sqrt{g}} \right] \hspace{0.1cm} .
\end{equation}
Here, the metric on the vortex sheet has been introduced,
\begin{equation}
g_{ab} (\sigma) = \partial_a \bar{x}_\mu (\sigma) \partial_b \bar{x}_{\mu}
(\sigma)  \hspace{0.1cm} , \hspace{0.3cm} g = \det (g_{ab}) \ .
\end{equation}
Furthermore, a partial integration has been performed in the last term of
equation (\ref{G4}). The expression obtained above can be cast into a more
standard form by using the decomposition
\begin{equation}
\partial_a \partial_b \bar{x}_\mu (\sigma ) = 
\Gamma^{c}_{ab} \partial_{c} \bar{x}_{\mu } +
K^A_{ab} n^A_{\mu } \ ,
\label{kabdec}
\end{equation}
where $\Gamma^{c}_{ab}$ are the affine connections and $K^A_{ab}$ denotes
the extrinsic curvature. Furthermore, $n^A_{\mu } $, with $A = 1,2$, are 
unit vectors orthogonal to the tangent vector of the vortex sheet
\begin{equation}
n^A_\mu \partial_c \bar{x}_\mu (\sigma) = 0 \hspace{0.1cm} , 
\hspace{0.1cm} n^A_\mu n^B_\mu =
\delta^{AB} \hspace{0.1cm} , \hspace{0.1cm} A = 1, 2 \hspace{0.1cm} .
\end{equation}
Using
\begin{equation}
\epsilon^{ab} g_{bd} \epsilon^{cd} = g^{ac} \cdot g
\end{equation}
one eventually obtains after straightforward evaluation
\begin{equation}
\bar{S} = \left( \frac{\Lambda }{\sqrt{2 \pi} } \right)^{2} 
\int_{\partial \Sigma } d^2 \sigma
\sqrt{g} \left[ 2 - \frac{1}{\Lambda^2} g^{ab} g^{ik} K^A_{ai} K^A_{bk}
+ \frac{\sqrt{g}}{2 \Lambda^2} g^{ab} \partial_a \partial_b 
\ln \sqrt{g} \right]
\hspace{0.1cm} .
\end{equation}
The first term, which diverges for $\Lambda \to \infty$, is the familiar
Nambu-Goto term. The remaining two terms are finite for $\Lambda \to \infty$.
The second term, containing the extrinsic curvature coefficients $K^a_{ai}$,
has been discussed in \cite{polyea},\cite{kleinea}. The higher-order 
terms in the gradient expansion all vanish for $\Lambda \to \infty$.
\medskip

\subsection{The thin center vortex}
\label{appthv}
In the following, it is shown that the thin center vortex as defined
by eq. (\ref{arbthv}) for arbitrary vortex shapes $S=\partial \Sigma $ 
indeed yields a center element for the Wilson loop when non-trivially linked
to it. For this purpose, it is convenient to write the thin vortex as
\begin{eqnarray}
a_{\mu } (k,S,x) &=& E(k) \alpha_{\mu } (S,x) \\
\alpha_{\mu } (S,x) &=& \int_{S} d^{D-2} \tilde{\sigma }_{\mu \kappa }
\partial_{\kappa }^{x} D(x-\bar{x} (\sigma ) )
\end{eqnarray}
Using Stokes' theorem, one obtains
\begin{eqnarray}
\oint_{C} dx_{\mu } \alpha_{\mu } (S,x) &=&
\int_{M(C)} d^2 \sigma_{\lambda \mu } \partial_{\lambda }^{x}
\alpha_{\mu } (S,x) \\
&=& \int_{M(C)} d^2 \sigma_{\lambda \mu }
\int_{S} d^{D-2} \tilde{\sigma }_{\mu \kappa }
\partial_{\lambda }^{x} \partial_{\kappa }^{x} D(x-\bar{x} (\sigma ) )
\end{eqnarray}
where $C=\partial M$. By using the relation between $d\sigma_{\mu \nu } $
and its dual $d\tilde{\sigma }_{\mu \nu } $, the last expression can be
rewritten as
\begin{equation}
\oint_{C} dx_{\mu } \alpha_{\mu } (S,x) = -\frac{1}{4} \int_{M(C)} 
d^2 \tilde{\sigma }_{\alpha^{\prime } \beta^{\prime } }
\int_{S} d^{D-2} \tilde{\sigma }_{\alpha \beta }
\epsilon_{\mu \nu \alpha^{\prime } \beta^{\prime } }
\epsilon_{\mu \kappa \alpha \beta }
\partial_{\lambda }^{x} \partial_{\kappa }^{x} D(x-\bar{x} (\sigma ) )
\end{equation}
Using the properties of $\epsilon_{\alpha \beta \gamma \delta } $, this 
expression can be converted to
\begin{eqnarray}
\oint_{C} dx_{\mu } \alpha_{\mu } (S,x) &=&
-\frac{1}{2} \int_{M(C)} d^{D-2} \tilde{\sigma }_{\alpha \beta } (x)
\int_{S} d^2 \sigma_{\alpha \beta } (\bar{x} ) \partial_{x}^{2}
D(x-\bar{x} ) \\ & & \ \ \ \ \ \ \ \ +
\int_{M(C)} d^{D-2} \tilde{\sigma }_{\kappa \beta }
\int_{S} d^2 \sigma_{\alpha \beta } \partial_{\alpha } \partial_{\kappa }
D(x-\bar{x} )
\end{eqnarray}
The last term is seen to vanish by Stokes' theorem, since
$\partial S = \partial (\partial \Sigma ) =0$. The first term yields, upon 
using the definition of the Green's function, the intersection number between
$M(C)$ and $S$,
\begin{eqnarray}
\oint_{C} dx_{\mu } \alpha_{\mu } (S,x) &=&
\frac{1}{2} \int_{M(C)} d^{D-2} \tilde{\sigma }_{\alpha \beta } (x)
\int_{S} d^2 \sigma_{\alpha \beta } (\bar{x} ) \delta^{D} (x-\bar{x} ) \\
&=& I(M(C),S) \\ &=& L(C,S)
\end{eqnarray}
which equals the linking number between $C$ and $S$, i.e. $L(C,S)$.
With the last relation, one indeed obtains a center element for the 
Wilson loop from the thin vortex (\ref{arbthv}),
\begin{equation}
\exp \left( -\oint_{C} dx_{\mu } a_{\mu } (k,S,x) \right)
= \exp (-E(k) L(C,S) ) = Z(k)^{L(C,S)}
\end{equation}

\section{Intersection number of two-dimensional \\ sheets in $D=4$}
\label{intseca}
This appendix summarizes essential properties of the self-intersection 
number of two-dimensional surfaces in $D=4$ space-time dimensions;
in particular, the relation of the self-intersection number to topological 
invariants of the three-dimensional spaces located at the initial and 
final times is presented. Assume a certain parametrization of the vortex 
sheet $\bar{x}_{\mu} (\sigma) \equiv \bar{x}_{\mu} (\sigma_1, \sigma_2)$ 
in $D = 4$. The self-intersection number $I(S,S)$, cf. eq. (\ref{isndef}),
i.e., the intersection number of $S$ with itself, receives 
contributions from ``coincidence points'' 
$\bar{x}_{\mu} (\sigma ) = \bar{x}_{\mu} (\sigma^{\prime } )$ with 
$\sigma = \sigma^{\prime } $ and from (transversal) ``intersection points'' 
$\bar{x}_{\mu} (\sigma) = \bar{x}_{\mu} (\sigma^{\prime } )$ with
$\sigma \neq \sigma^{\prime } $, which will be denoted by $I_1 (S,S)$ and 
$I_2 (S,S)$ respectively,
\begin{eqnarray}
I(S,S) = I_1 (S,S) + I_2 (S,S)
\end{eqnarray}
Regularizing the delta function in eq. (\ref{isndef}), cf. Appendix 
\ref{idappv}, eq. (\ref{regidv}), introduces an ultraviolet cutoff 
$\Lambda $, which then allows $I_1 (S,S)$ to be straightforwardly evaluated; 
in the limit $\Lambda \rightarrow \infty $, one finds Polyakov's 
intersection number\footnote{$I_1 (S,S)$ differs from Polyakov's original 
definition \cite{polyea} by a factor of $\left( -\frac{1}{4} \right) $.}, 
\begin{eqnarray}
I_1 (S,S) = - \frac{1}{16\pi } \int_{S} \frac{d^2 \sigma }{\sqrt{g} }
g^{ij} \partial_{i} \Sigma_{\mu \nu } \partial_{j} 
\tilde{\Sigma }_{\mu \nu }
\end{eqnarray}
where
\begin{eqnarray}
g_{ij} = \partial_{i} \bar{x}_{\mu } (\sigma ) \partial_{j} 
\bar{x}_{\mu } (\sigma )
\end{eqnarray}
denotes the metric on the surface $S$, $g = \det g_{ij} $ and 
\begin{eqnarray}
\Sigma_{\mu \nu } = \epsilon_{ij} \partial_{i} \bar{x}_{\mu } (\sigma ) 
\partial_{j} \bar{x}_{\nu } (\sigma )
\end{eqnarray}
is the tensor of the surface element on the vortex sheet.
\medskip

Let $\bar{x}_{\mu } (\sigma (k)) = \bar{x}_{\mu } (\sigma^{\prime } (k))$,
where $\sigma (k) \neq \sigma^{\prime } (k)$, determine the coordinates of 
isolated (self-)intersection points $P(k)$ of $S$. Expanding
$\bar{x}_{\mu } (\sigma ), \bar{x}_{\mu } (\sigma^{\prime } )$ near 
$\sigma = \sigma (k), \sigma^{\prime } = \sigma^{\prime } (k)$ in powers 
of $\sigma - \sigma (k), \sigma^{\prime } - \sigma^{\prime } (k)$,
the integrals in (\ref{isndef}) over $d^2 \sigma , d^2 \sigma^{\prime } $ 
can be straightforwardly performed, yielding
\begin{equation}
I_2 (S,S) = 2 \sum_{k} \mbox{sign} \, \left( \Sigma_{\mu \nu } 
(\bar{x} (\sigma (k))) \tilde{\Sigma }_{\mu \nu } (\bar{x} (\sigma (k)))
\right)
\end{equation}
One observes that each intersection point $\bar{x}_{\mu } (\sigma (k)) = 
\bar{x}_{\mu } (\sigma^{\prime } (k))$, with
$\sigma (k) \neq \sigma^{\prime } (k)$, contributes doubly to the 
self-intersection number $I_2 (S,S)$ in agreement with the findings in
section \ref{topsect}.
\medskip

To illustrate the geometric meaning of the different contributions to the
intersection number, it is useful to consider a fixed time slice of the
four-dimensional universe. In this three-dimensional space, the vortex 
$S$ represents a closed loop $C (t = \mbox{fixed} )$. For simplicity,
choose $t = \sigma_{2} $ such that this loop is parametrized by 
$\sigma_{1} $, which takes values in $[0, 2\pi [$.
\medskip

Consider the self-linking number $SL(C, \hat{n} )$ of such a closed loop 
$C(t)$ in $D = 3$, which is defined by the linking number $L(C,C^{\prime } )$
between the original loop $C(t)$ and the fictitious loop $C^{\prime } (t)$ 
resulting from $C(t)$ by an infinitesimal shift 
\begin{eqnarray}
\bar{x}_{\mu } (\sigma_{1} , t) \rightarrow \bar{x}_{\mu } (\sigma_{1}, t) 
+ \epsilon \hat{n} (\sigma_{1} , t) \ .
\end{eqnarray}
Here, $\hat{n} (\sigma_{1} ,t)$ is a unit vector perpendicular to the 
tangent vector $\vec{\bar{x} }^{\prime } (\sigma_{1} , t) = 
\frac{\partial}{\partial \sigma_{1} } \vec{\bar{x} } (\sigma_{1} ,t)$ 
of the loop $C(t)$,
\begin{eqnarray}
\hat{n} (\sigma_{1} ,t) \cdot \vec{\bar{x} }^{\prime } (\sigma_{1} , t) 
= 0 \, , \ \ \ \hat{n}^{2} = 1
\end{eqnarray}
which is periodic in $\sigma_{1} , \hat{n} (\sigma_{1} + 2\pi , t) =
\hat{n} (\sigma_{1} )$, i.e.
\begin{equation}
SL(C, \hat{n} ) = \lim_{\epsilon \rightarrow 0} L(C,C^{\prime } ) \ .
\end{equation}
In this last equation,
\begin{eqnarray}
L(C,C^{\prime } ) = \frac{1}{4\pi } \oint_{C} \oint_{C^{\prime } } 
d\vec{x} \times d\vec{x}^{\prime }
\frac{\vec{x} - \vec{x}^{\prime } }{| \vec{x} - \vec{x}^{\prime } |}
\end{eqnarray}
is the familiar Gau\ss \ linking number. 
\medskip

The vector $\hat{n} (\sigma_{1} , t)$ defines a framing of the loop $C(t)$,
which implies a thickening of the loop into a ribbon bounded by $C(t)$ and 
$C^{\prime } (t)$. Let $\bar{C} $ denote the loop along the middle of this 
ribbon which is generated by $\vec{\bar{x} } (\sigma_{1} , t) + 
\frac{\epsilon }{2} \hat{n} (\sigma_{1} , t)$ with $t$ fixed. Taking the 
limit $\epsilon \rightarrow 0$, one finds for the self-linking
number \cite{kleinbu}
\begin{eqnarray}
SL(C, \hat{n} ) = W_{r} (C) + T_{w} (C, \hat{n} )
\end{eqnarray}
Here 
\begin{eqnarray}
W_{r} (C) = L(\bar{C} ,\bar{C} )
\end{eqnarray}
is the writhing number and 
\begin{eqnarray}
T_{w} (C,\hat{n} ) = \frac{1}{2\pi } \oint d\sigma_{1} 
\left[ \vec{e} (\sigma_{1} , t) \times \hat{n} (\sigma_{1} , t) \right] 
\cdot \hat{n}^{\prime } (\sigma_{1} , t)
\end{eqnarray}
denotes the twist, where 
\begin{eqnarray}
\vec{e} (\sigma_{1} , t) = 
\frac{\vec{\bar{x} }^{\prime } (\sigma_{1} , t)}{| \vec{\bar{x} }^{\prime } 
(\sigma_{1}, t) |}
\end{eqnarray}
is the unit vector tangential to the loop $C(t)$. The twist 
$T_{w} (C, \hat{n} )$ is not a topological invariant but depends on 
the choice of the framing $\hat{n} (\sigma_{1} , t)$. A global change of 
$ \hat{n} (\sigma_{1} , t)$ changes $T_{w} (C,\hat{n} )$ by some 
integer. The twist $T_{w} (C, \hat{n} )$ expresses the torsion of the 
ribbon bounded by $C(t)$ and $C^{\prime } (t)$ and represents the Polyakov 
spin factor in $D = 3$. It should be emphasized that, while the 
self-linking number and the twist depend on the choice of the framing 
$\hat{n} (\sigma_{1} , t)$, their difference, i.e. the writhing number, 
is independent of this arbitrariness. Furthermore, it can be shown that the 
self-intersection number $I_2 (S,S)$ is related to the difference in the 
self-linking number between the initial and final time $t_1 $ and $t_2 $ of
the four-dimensional space-time manifold \cite{satoor}
\begin{eqnarray}
I_2 (S,S) = \left. - SL (C(t), \hat{n} (t)) \right|^{t=t_2 }_{t=t_1 } \ .
\end{eqnarray}
In an analogous way, Polyakov's intersection number is related to the twist 
of $C(t)$,
\begin{equation}
I_1 (S,S) = \left. T_{w} (C(t), \hat{n} (t)) \right|^{t=t_2 }_{t=t_1 } \ ,
\end{equation}
so that the total intersection number is given by the writhing number 
\begin{equation}
I (S,S) = \left. - W_{r} (C(t)) \right|^{t=t_2 }_{t=t_1 } \ .
\end{equation}
A transversal self-intersection point of the vortex sheet gives rise to a
self-intersection point of the vortex string $C(t)$ at some time $\bar{t} $. 
If the loop $C(t)$ self-intersects at some time $t = \bar{t} $, the 
writhing number changes by 2 from the time before $\bar{t} $ until the time 
after the intersection. This contribution 2 corresponds to the fact that 
each (transversal) self-intersection point of the vortex sheet contributes 
2 units to the intersection number. Since for all closed vortex sheets,
self-intersection points occur pairwise, the vortex string $C(t)$ must 
self-intersect an even number of times, so that the $SU(2)$ Pontryagin 
index 
\begin{equation}
\nu [ A(S) ] = 
\left. - \frac{1}{4} W_{r} (C(t)) \right|^{t=t_2 }_{t=t_1 }
\end{equation}
becomes an integer, as observed already in section \ref{topsect}.
\medskip

\section{Topological charge of a thick SU(2) vortex intersection}
\label{appc}
Consider two intersecting planar thick $SU(2)$ vortices in four 
space-time dimensions. By choice of coordinate system, the intersection
region can be centered on the origin of space-time, and one of the vortices 
can be centered on the 1-2-plane,
\begin{equation}
A^{(1)} = T_3 \frac{f_1 (x_3^2 +x_4^2 )}{x_3^2 +x_4^2 } (0,0,x_4 ,-x_3 )
\end{equation}
cf. eq. (\ref{thickv}), with an arbitrary profile function $f_1 $.
On the other hand, let the other vortex be centered on the
$3^{\prime }-4^{\prime } $-plane of a (primed) coordinate system related 
to the previous one by an arbitrary orthogonal transformation $O$,
\begin{eqnarray}
A^{(2) \prime } &=& T_3 
\frac{f_2 (x_1^{\prime 2} +x_2^{\prime 2} )}{x_1^{\prime 2} 
+x_2^{\prime 2} } (x_2^{\prime } , -x_1^{\prime } ,0,0) \\
x_i^{\prime } &=& O_{ij} x_j
\end{eqnarray}
This also defines the transformation of the direction of the four-vector 
$A^{(2) \prime } $ when going to the unprimed coordinate system,
$A^{(2)} = O^{T} A^{(2) \prime } $. In the unprimed coordinate system, 
therefore, the second thick vortex is given by
\begin{equation}
A^{(2)}_{i} = T_3 \frac{f_2 [ (O_{1j} x_j )^2 
+ (O_{2k} x_k )^2 ]}{(O_{1l} x_l )^2 + (O_{2m} x_m )^2 }
\left( O_{1i} O_{2p} x_p - O_{2i} O_{1q} x_q \right)
\end{equation}
From this, one straightforwardly obtains the field strengths
associated with the thick vortex fields (where the prime on the
profile functions denotes differentiation w.r.t. the argument),
\begin{eqnarray}
F^{(1)}_{34} = -F^{(1)}_{43} &=& 
\partial_{3} A^{(1)}_{4} - \partial_{4} A^{(1)}_{3} 
= -2T_3 f_1^{\prime } (x_3^2 + x_4^2 ) \\
F^{(1)}_{ij} &=& 0 \ \ \ \mbox{otherwise}
\end{eqnarray}
and
\begin{eqnarray}
F^{(2)}_{ij} &=& \partial_{i} A^{(2)}_{j} - \partial_{j} A^{(2)}_{i} \\
&=& 2T_3 \left( O_{1j} O_{2i} - O_{2j} O_{1i} \right)
f_2^{\prime } \left( (O_{1k} x_k )^2 + (O_{2l} x_l )^2 \right) \ .
\end{eqnarray}
The Pontryagin index is given by
\begin{equation}
\nu = -\frac{1}{32\pi^{2} } \int d^4 x \, \epsilon_{ijkl} 
\mbox{tr} (F_{ij} F_{kl} )
\end{equation}
One can easily verify that the integrand vanishes for each thick
vortex taken separately,
\begin{equation}
\epsilon_{ijkl} \mbox{tr} (F^{(1)}_{ij} F^{(1)}_{kl} ) =
\epsilon_{ijkl} \mbox{tr} (F^{(2)}_{ij} F^{(2)}_{kl} ) = 0
\end{equation}
and therefore the only nonvanishing contribution to $\nu $ results from
the intersection region of the two thick vortices,
\begin{eqnarray}
\nu &=& -\frac{1}{4\pi^{2} } \int d^4 x \, 
\mbox{tr} (F^{(2)}_{12} F^{(1)}_{34} ) \\
&=& -\frac{1}{2\pi^{2} } \int d^4 x \,
(O_{12} O_{21} - O_{22} O_{11} ) \, \,
f_2^{\prime } \left( (O_{1k} x_k )^2 + (O_{2l} x_l )^2 \right)
f_1^{\prime } (x_3^2 + x_4^2 ) \nonumber
\end{eqnarray}
By the change of integration variables
\begin{equation}
(x_1 ,x_2 ) \longrightarrow 
(x_1^{\prime } , x_2^{\prime } ) = (O_{1k} x_k , O_{2l} x_l )
\end{equation}
this simplifies (up to the sign of the Jacobian) to
\begin{eqnarray}
| \nu | &=& \frac{1}{2\pi^{2} } \int dx_1^{\prime } dx_2^{\prime } dx_3 dx_4
f_2^{\prime } (x_1^{\prime 2} + x_2^{\prime 2} )
f_1^{\prime } (x_3^2 + x_4^2 ) \\
&=& 2\int_{0}^{\infty } dr \int_{0}^{\infty } dr^{\prime }
r r^{\prime } f_2^{\prime } (r^{\prime 2} ) f_1^{\prime } (r^2 ) \\
&=& \frac{1}{2} \left. f_2 (r^{\prime 2} ) \right|_{0}^{\infty }
\cdot \left. f_1 (r^2 ) \right|_{0}^{\infty }
\end{eqnarray}
Due to the boundary conditions on profile functions (cf. the description
after eq. (\ref{thickv})), one thus finally obtains
\begin{equation}
| \nu | = \frac{1}{2} \ ,
\end{equation}
generalizing the result for thin $SU(2)$ vortices derived in section
\ref{topsect}. The specific form of the profile functions has completely
canceled, and the present treatment thus in particular also
includes the case of thin $SU(2)$ vortices. Also all reference to the
coordinate rotation $O$ specifying the relative orientation of the
vortex planes has vanished.

\end{appendix}

\end{document}